\documentclass[journal,11pt]{IEEEtran}

\usepackage{amsmath}
\usepackage{amsfonts}
\usepackage{mathtools}
\usepackage{epsfig}
\usepackage{calrsfs}
\usepackage{ifthen}
\usepackage{cancel} 
\usepackage{multirow}
\usepackage{colortbl}
\usepackage{adjustbox}
\usepackage{lscape}
\usepackage{tikz}
\usetikzlibrary{positioning}
\usepackage{colortbl}
\usepackage{paralist}
\usepackage{algorithm}
\usepackage{algorithmic}
\usepackage{multicol}
\usepackage{extarrows}
\usepackage{tabularx}
\usepackage{hhline}
\usepackage{wrapfig}
\usepackage{graphicx}
\usepackage{subcaption}
\usepackage{array}
\newcolumntype{L}[1]{>{\raggedright\let\newline\\\arraybackslash\hspace{0pt}}m{#1}}
\newcolumntype{C}[1]{>{\centering\let\newline\\\arraybackslash\hspace{0pt}}m{#1}}
\newcolumntype{R}[1]{>{\raggedleft\let\newline\\\arraybackslash\hspace{0pt}}m{#1}}
\usepackage{verbatimbox}
\usepackage{bmpsize}
\usepackage{xcolor}
\usepackage{lipsum}
\usepackage[normalem]{ulem}
\usepackage{flushend}

\RequirePackage{expl3}
\ExplSyntaxOn
\seq_new:N \l_pclist_classes_seq
\seq_new:N \l_pclist_packages_seq
\seq_new:N \l_pclist_other_seq
\clist_map_inline:cn { @filelist }
  {
   \tl_if_in:nnTF { #1 } { .cls }
     {
      \tl_set:Nn \l_tmpa_tl { #1 }
      \tl_remove_once:Nn \l_tmpa_tl { .cls }
      \seq_put_right:NV \l_pclist_classes_seq \l_tmpa_tl
     }
     {
      \tl_if_in:nnTF { #1 } { .sty }
       {
        \tl_set:Nn \l_tmpa_tl { #1 }
        \tl_remove_once:Nn \l_tmpa_tl { .sty }
        \seq_put_right:NV \l_pclist_packages_seq \l_tmpa_tl
       }
       {
        \seq_put_right:Nn \l_pclist_other_seq { #1 }
       }
     }
  }

\newboolean{ieee}
\newboolean{lncs}
\newboolean{sig}
\newboolean{acm}
\newboolean{usenix}
\newboolean{elsarticle}

\seq_if_in:NnTF \l_pclist_classes_seq { article }
  { \setboolean{usenix}{true} }
  { \setboolean{usenix}{false} }
\seq_if_in:NnTF \l_pclist_classes_seq { IEEEtran }
  { \setboolean{ieee}{true} }
  { \setboolean{ieee}{false} }
\seq_if_in:NnTF \l_pclist_classes_seq { llncs }
  { \setboolean{lncs}{true} }
  { \setboolean{lncs}{false} }
\seq_if_in:NnTF \l_pclist_classes_seq { sig-alternate }
  { \setboolean{sig}{true} }
  { \setboolean{sig}{false} }
\seq_if_in:NnTF \l_pclist_classes_seq { acmart }
  { \setboolean{acm}{true} }
  { \setboolean{acm}{false} }
\seq_if_in:NnTF \l_pclist_classes_seq { elsarticle }
  { \setboolean{elsarticle}{true} }
  { \setboolean{elsarticle}{false} }
\seq_if_in:NnTF \l_pclist_classes_seq { usenix }
  { \setboolean{usenix}{true} }
  { \setboolean{usenix}{false} }

\ExplSyntaxOff

\newboolean{xl} 
\setboolean{xl}{false} 
\ifthenelse{\boolean{sig}}{
  \usepackage[hyphens]{url} 
  \usepackage[ breaklinks, colorlinks=true, linkcolor=black]{hyperref} 
}{
  \usepackage{amsthm} 
}

\ifthenelse{\boolean{acm}}{
}{
  \usepackage{amssymb} 
}

\ifthenelse{\boolean{ieee}}{
  \usepackage[hyphens]{url} 
  \usepackage[ breaklinks, colorlinks=true, urlcolor=blue, linkcolor=black]{hyperref}
}{}

\ifthenelse{\boolean{lncs}}{
  \usepackage[ breaklinks, colorlinks=true, linkcolor=black]{hyperref}
  \usepackage[hyphens]{url} 
}{}

\ifthenelse{\boolean{elsarticle}}{
  \usepackage{hyperref}
  \usepackage{url} 
}{}

\newboolean{showdelete} 
\setboolean{showdelete}{false}

\usepackage[moderate,mathspacing=normal,bibliography=normal, margins=normal]{savetrees}

\newcommand{\subhead}[1]{\vspace {1pt}\noindent{\textbf{#1.}}}

\newcommand{\lurk}{LURK-T}

\newcommand{\versionone}{DHE-active \cs{}}
\newcommand{\versiontwo}{DHE-passive \cs{}}
\newcommand{\dhea}{``\lurk\ with \versionone''}
\newcommand{\dhep}{``\lurk\ with \versiontwo''}

\newcommand{\keyless}{keyless}

\newcommand{\party}[1]{\ensuremath{\mathit{#1}}}  
\newcommand{\client}{\party{C}}
\newcommand{\cs}{\party{CS}}
\newcommand{\engine}{\party{E}}
\newcommand{\server}{\party{S}}
\newcommand{\attester}{\party{AS}}

\newcommand{\cnonce}{\ensuremath{\msg{N}_{\client}}}
\newcommand{\snonce}{\ensuremath{\msg{N}_{\server}}}
\newcommand{\enonce}{\ensuremath{\msg{N}_{\engine}}}

\newcommand{\msg}[1]{\ensuremath{\mathsf{#1}}} 
\newcommand{\ckeyshare}{\ensuremath{\msg{KE}_{\client}}}
\newcommand{\skeyshare}{\ensuremath{\msg{KE}_{\server}}}
\newcommand{\ecdhesharesecret}{\ensuremath{\msg{KE}}}
\newcommand{\handshake}{\ensuremath{\msg{H}_{ctx}}}
\newcommand{\ssign}{\ensuremath{\mathsf{PSign}}}

\newcommand{\certv}{\ensuremath{\msg{CertificateVerify}}}

\newcommand{\sfin}{\ensuremath{\msg{Fin}_{\engine}}}
\newcommand{\cfin}{\ensuremath{\msg{Fin}_{\client}}}

\newcommand{\alg}[1]{\ensuremath{\mathtt{#1}}} 
\newcommand{\algae}{\ensuremath{\alg{AE}}}
\usepackage{xparse}
\NewDocumentCommand\am{g}{%
  \IfNoValueF{#1}{{\color{blue} \textbf{(AM: #1)}}}%
  \IfNoValueT{#1}{{\color{blue} \textbf{(AM)}}}%
}

\newcommand{\mycvf}[1]{\ensuremath{\msg{CVf}_{#1}}} %

    \newcommand{\scvf}[1]{\mycvf{\myserver}}

\newcommand{\myserver}{\party{S}}

\newcommand{\ck}{\msg{ck}}

\newcommand{\adv}{\ensuremath{\mathcal{A}}}

\newcommand\copyrighttext{%
  \footnotesize \textcopyright 2024 IEEE. Personal use of this material is permitted.
  Permission from IEEE must be obtained for all other uses, in any current or future
  media, including reprinting/republishing this material for advertising or promotional
  purposes, creating new collective works, for resale or redistribution to servers or
  lists, or reuse of any copyrighted component of this work in other works. }
\newcommand\copyrightnotice{%
\begin{tikzpicture}[remember picture,overlay]
	\node[anchor=south,yshift=6pt] at (current page.south) {\fbox{\parbox{\dimexpr\textwidth-\fboxsep-\fboxrule\relax}{\copyrighttext}}};
\end{tikzpicture}%
}

\begin{document}


\ifthenelse{\boolean{elsarticle}}{

\begin{frontmatter}

\title{\lurk{}:  Limited Use of Remote Keys with Added Trust in TLS 1.3}
\author[inst1]{Behnam Shobiri}
\ead{behnam.shobiri@concordia.ca}
\author[inst1]{Sajjad Pourali}
\ead{s\_poural@ciise.concordia.ca}
\author[inst2]{Daniel Migault}
\ead{daniel.migault@ericsson.com}
\author[inst3]{Ioana Boureanu}
\ead{i.boureanu@surrey.ac.uk}
\author[ins2]{Stere Preda}
\ead{stere.preda@ericsson.com}
\author[inst1]{Mohammad Mannan\corref{cor1}}
\ead{m.mannan@concordia.ca}
\author[inst1]{Amr Youssef}
\ead{youssef@ciise.concordia.ca}

\affiliation[inst1]{organization={Concordia University},
            city={Montreal},
            state={Quebec},
            country={Canada}}

\affiliation[inst2]{organization={Ericsson},
            city={Montreal},
            state={Quebec},
            country={Canada}}

\affiliation[inst3]{organization={University of Surrey},
city={ Guildfordl},
country={UK}}

\cortext[cor1]{Corresponding author}

\end{frontmatter}
}{}

\ifthenelse{\boolean{ieee}}{
\title{\lurk{}:  Limited Use of Remote Keys\\With Added Trust in TLS 1.3}

\author{Behnam~Shobiri, 
Sajjad~Pourali,
Daniel~Migault,
Ioana~Boureanu,
Stere~Preda,
Mohammad~Mannan\\
and~Amr~Youssef,~\IEEEmembership{Senior Member,~IEEE}
\thanks{
Manuscript received 10 March 2023; revised 22 August 2023; accepted
19 July 2024. Date of publication 23 July 2024; date of current version
15 November 2024. This work was supported by Mitacs Accelerate Cluster
through Ericsson Research Canada. Recommended for acceptance by Dr. Pan
Zhou. (Corresponding author: Mohammad Mannan.)\\
Daniel Migault and Stere Preda are with the Ericsson, Montreal, QC H4S 0B6,
Canada (e-mail: daniel.migault@ericsson.com; stere.preda@ericsson.com).\\
Ioana Boureanu is with the Surrey Centre for Cyber Security, the Department
of Computer Science, University of Surrey, GU2 7XH Guildfordl, U.K. (e-mail:
i.boureanu@surrey.ac.uk).\\
Digital Object Identifier 10.1109/TNSE.2024.3432836
%

}}

\markboth{IEEE Transactions on Network Science and Engineering,~Vol.~11, No.~6, NOVEMBER/DECEMBER~2024}%
{Shell \MakeLowercase{\textit{et al.}}: Bare Demo of IEEEtran.cls for IEEE Journals}

\maketitle
\copyrightnotice

\begin{abstract}
In many web applications, such as Content Delivery Networks (CDNs),  TLS credentials are shared, e.g.,  between 
the website’s TLS origin server  and the CDN's  edge servers, which can be distributed around the globe.
To  enhance the security and trust for TLS~1.3 in such scenarios, we propose {\lurk{}}, a provably secure framework which allows for  limited use of remote keys with added trust in TLS 1.3. 
We efficiently decouple the server side of TLS~1.3 into a {\lurk{} Crypto Service (\cs{})} and a {\lurk{}  Engine (\engine{})}. \cs{} executes all cryptographic operations in a Trusted Execution Environment (TEE), upon \engine{}'s requests.  \cs{} and \engine{} together provide the whole TLS-server functionality. 
A major benefit of our construction is that it is application agnostic; the {\lurk{} Crypto Service}  could be collocated with the {\lurk{}  Engine}, or it could run on different machines. Thus, our  design  allows for in situ attestation and protection of the cryptographic side of the TLS server, as well as for all setups of CDNs over TLS. 
To support such a generic decoupling, we provide a full  Application Programming Interface (API) for \lurk{}.
To this end, we  implement our \lurk{} Crypto Service using Intel SGX and integrate it with OpenSSL.
We also test \lurk's efficiency and show that, from a TLS-client's perspective, HTTPS servers using \lurk{} instead a traditional TLS-server have no noticeable 
overhead 
when serving files greater than 1MB.
In addition, we provide  cryptographic proofs and formal security verification using ProVerif.

\end{abstract}

}{}


\section{Introduction}
\label{sec:intro}

Transport Layer Security (TLS) is  the de-facto   protocol for securing communication over the Internet. It is an authenticated key-establishment (AKE) protocol, whereby TLS client \client\ (e.g., browser) always authenticates a TLS server \server, and they derive \emph{channel keys} to communicate securely thereafter. 
In TLS, the server \server{} is authenticated by proving the possession of its private key or a so-called pre-shared key (PSK). So, these authentication credentials should not be accessible by other parties and require special attention. 

For TLS servers managed  ``in situ'', e.g., when the owner of the TLS server also owns the infrastructure and entirely manage the TLS servers, the authentication credentials must be protected for example against operational mistakes\footnote{\url{https://lists.dns-oarc.net/pipermail/dns-operations/2020-May/020198.html}}
as well as web server compromise  such as Heartbleed.\footnote{\url{https://heartbleed.com/}}

With 73\% of the Internet traffic today being served by Content Delivery Networks (CDNs)~\cite{cisco2017cisco},
a common scenario is  sharing the TLS credentials between the website's TLS server (i.e., the ``origin'') and the CDN's ``edge servers'', 
which can be distributed around the globe. Such sharing of long-term TLS credentials poses a grave risk, as the origin  loses full ownership and  control of their long-term private key~\cite{DBLP:conf/sp/LiangJDLWW14}.  

The proposed setups for CDN over TLS alone vary vastly
from splitting the TLS implementation~\cite{DBLP:conf/trustcom/StebilaS15}, to leveraging Trusted Execution Environment (TEE) and either improving the performances of the enclaves for network applications~\cite{DBLP:conf/isca/WeisseBA17, DBLP:conf/uss/HerwigGL20, DBLP:conf/ndss/ShindeTTS17} or improving a specific application running inside an enclave~\cite{talos} -- which ends into splitting the application between components running inside the TEE and outside the TEE.
Thus,  a  generic treatment of securing and protecting the long-term credentials of the TLS server is essential, catering for as many distinct types of interactions as possible. To this end, we propose \lurk{}: a generic, provably secure and efficient decoupling of the TLS1.3 server into a cryptographic core  called \emph{\lurk\ Crypto Service (\cs{})}, and a component called \emph{\lurk\ Engine (\engine{})} which securely queries this core from anywhere it may reside,  and communicates  with a classical TLS Client (\client{}).   


We are not the first to consider the decoupling of a TLS server and/or securing a modified version thereof. Current efforts can be divided into two types: (a)  TEE-driven approach focusing on isolating and securing the server; (b) CDN-driven approach focusing on modifying the TLS server to fit different CDN setups. Each approach has its merits and  shortcomings. Inspired by both these approaches, we propose a new solution, 
by decoupling the TLS-server in a way that results into acceptable, deployment-friendly performance.
Now, we discuss the two main aspects of our design compared to existing work (details in Section~\ref{sec:relW}). 

\textbf{(a) TLS servers and CDNs.} CDNs operate over TLS in a mechanism often broadly referred to as ``TLS delegation''. To enable such delegation in a provably secure way (as in e.g.,~\cite{DBLP:conf/trustcom/BoureanuMPAMFM20}), or to support specific scenarios~\cite{draft-ietf-tls-subcerts},
major operational changes in TLS are required. Such changes either break security (see e.g.,~\cite{DBLP:conf/trustcom/StebilaS15}), 
or render them completely incompatible with legacy clients (see e.g.,~\cite{DBLP:conf/sigcomm/NaylorSVLBLPRS15}).
Besides, the efficiency of delegation is usually not  considered/discussed at length or is sacrificed in favor of enhanced security (see e.g.,~\cite{SAndP18}).

\textbf{(b) TLS servers and TEEs.} To protect  TLS credentials,  
NIST~\cite{NISTIR8320} 
recommends hardware-based TEEs such as Trusted Platform Modules (TPMs) or Hardware Security Modules (HSMs), for storing and using private keys. 
Yet, due to significant cost and performance issues of large-scale HSM deployment, 
such TEE integration is not common for CDN scenarios. 
TEE-based academic proposals vary significantly where 
the full application is placed in a TEE~\cite{DBLP:conf/isca/WeisseBA17, DBLP:conf/uss/HerwigGL20,
DBLP:conf/ndss/ShindeTTS17, DBLP:conf/middleware/WeichbrodtAK18}, or the full TLS is placed in a TEE~\cite{talos} -- both of which are explicitly mentioned as impractical by several standard bodies such as ETSI,\footnote{\url{https://www.etsi.org/deliver/etsi_gr/NFV-SEC/001_099/009/01.02.01_60/gr_NFV-SEC009v010201p.pdf}} 
3GPP-SA3,\footnote{\url{https://www.3gpp.org/ftp/Specs/archive/33_series/33.848/33848-0c0.zip}} 
and ENISA.\footnote{\url{https://www.enisa.europa.eu/publications/nfv-security-in-5g-challenges-and-best-practices/@@download/fullReport}} 
Some other proposals protect only the keys~\cite{
DBLP:conf/uss/HerwigGL20,brandao2021hardening, DBLP:conf/cloud/WeiLLYG17}.
Indeed, deciding which part of the cryptographic side of TLS-server to include in a TEE, such as to yield added security without high performance penalty, appears to be non-trivial.

\noindent\textbf{Our contributions} can be summarized as follows:

\textbf{1.} To  enhance the security and trust for TLS~1.3 in applications where the TLS credentials are shared (e.g., in CDN applications), we propose \emph{Limited Use of Remote Keys with Added Trust (\lurk{})}.
To balance security and efficiency,  \lurk{} splits the TLS~1.3 server into two parts: a \emph{\lurk{}  Engine} (\engine{}) and \emph{\lurk{} Crypto Service} (\cs{}). 
\cs{} resides inside a TEE, and is only involved during the TLS handshake. \cs{} handles and ensures the confidentiality of  TLS-server credentials intrinsically needed for TLS key-security: private keys, PSK for session resumption, Elliptic Curve Ephemeral Diffie Hellman ((EC)DHE) keys to ensure Perfect Forward Secrecy (PFS). \engine{} handles the rest of server-side TLS.
Moreover, our design is such that \engine's queries to \cs{} cannot be made outside the scope of a fresh TLS~1.3 Key EXchange (KEX). See Figure~\ref{fig:lurk-overview} in Section~\ref{sec:our-protocol} for an overview of \lurk{} components.

\textbf{2.}  
We implement \cs{} using Intel SGX and integrate it with OpenSSL, both for Ubuntu and Windows. The  modularity of our design entails only localized changes to OpenSSL.
To show the compatibility and portability of our implementation, we develop a Rust HTTPS server and link it to our modified OpenSSL.

\textbf{3.} We test \lurk{}'s efficiency extensively, measuring different overheads compared to a standard TLS~1.3 handshake---for all the TLS~1.3 cipher suites and various \cs{} configurations.
We measure the maximum number of files served per second with HTTPS and show that in the worst case configuration, the client's overhead associated to \lurk{} is negligible for files equal or greater than 1MB. 
The server's overhead is limited to the TLS handshake and we measured it between 1.2\% and 33\% which is far less than similar solutions (see Section~\ref{sec:relW} and Table~\ref{tab:relw_perf}).

\textbf{4.} We present cryptographic proofs for \lurk{}, in a cryptographic model for multi-party TLS~\cite{DBLP:conf/eurosp/BhargavanBFOR17}, showing that \lurk{} provides three-party TLS security (\engine{}, \cs{}, and \client{}).
We also formally verify  \lurk's security using ProVerif, by first lifting the existing ProVerif specifications~\cite{prosecco,SP2017} of a pre-standard TLS~1.3 to a ProVerif model for the standard TLS~1.3~\cite{RFC8446}, and then proving TLS~1.3 security for  \lurk; thus, we show that \lurk{}  suffers no degradation in security compared to TLS~1.3, including attaining perfect forward secrecy.  
We achieve strong security guarantees (e.g., the accountability of~\cite{SAndP18}),  as well as add a new property of trust which we call  ``trusted key-binding'', achieved through 
the attestation of our TEE-based \cs{}.

\section{Related Work}
\label{sec:relW}

\lurk{} partitions TLS~1.3 into two independent micro services (\engine{} and \cs{}) with \cs{} hosted by a TEE.
In this section, we summarize related work on partitioning applications, as well as as protocol extensions that support TLS delegation, and multi-party TLS. 

\subsection{TLS and TEE}

Multiple frameworks are able to host unmodified binary code into a TEE enclave (see e.g.,~\cite{DBLP:journals/corr/abs-2109-01923}). These frameworks rely on libOS (e.g., Graphene~\cite{DBLP:conf/usenix/TsaiPV17}, 
SGX-LKL~\cite{DBLP:journals/corr/abs-1908-11143}), or musl-libc (e.g., SCONE~\cite{DBLP:conf/osdi/ArnautovTGKMPLM16}). However, this results in a large trusted code base (TCB)~\cite{DBLP:conf/codaspy/TianCHTB19} with a vast number of Line of Code (LoC) prone to bugs~\cite{DBLP:conf/uss/CloostersRD20} (and Iago attacks~\cite{DBLP:conf/asplos/CheckowayS13}), and with large overhead due to multiple ECALLs/OCALLs~\cite{DBLP:journals/access/SuzakiNOT21}. 

Partitioning applications is expected to address these drawbacks.
Specific manual approaches have been proposed for TLS~1.2 as in~\cite{talos}.
A more generic approach, based on marking  sensitive data in the source code for C/C++ applications has been proposed in \emph{Glamdring}~\cite{DBLP:conf/usenix/LindPMOAKRGEKFP17} and the execution of the resulting trusted part can be instrumented by sgx-perf~\cite{DBLP:conf/middleware/WeichbrodtAK18}. 
Other proposals such as 
Montsalvat~\cite{DBLP:conf/middleware/YuhalaMFST0GL21} partition Java code based on its byte-code.  However, the coexistence of the trusted and untrusted part is handled via remote procedure call (RPC)-like mechanisms, exposing the interface to Iago-like attacks, while providing little assurance that data or states are not leaked.
\lurk{} defines standard interfaces in~\cite{draft-mglt-lurk-tls13}, thus protecting \cs{} against Iago-like attacks while enabling remote execution of the \cs{}. 
The combination of \cs{} and \engine{} is also formally proven to not alter TLS~1.3 security 
following~\cite{DBLP:conf/eurosp/BhargavanBFOR17}, which showed that the lack of such formal verification can hide the existence of vulnerabilities 
(e.g., in Keyless SSL~\cite{DBLP:conf/trustcom/StebilaS15} and mcTLS~\cite{DBLP:conf/sigcomm/NaylorSVLBLPRS15}).

Various efforts (e.g.,~\cite{talos, DBLP:conf/cloud/WeiLLYG17, DBLP:conf/middleware/WeichbrodtAK18, DBLP:conf/uss/HerwigGL20, brandao2021hardening, DBLP:conf/ndss/ShindeTTS17, DBLP:conf/codaspy/TianCHTB19}) were made to leverage TEE, and port TLS applications into SGX enclaves. 
All these proposals were focused on TLS~1.2, and generally they place the full TLS stack into the TEE (e.g., TaLoS~\cite{talos} and sgx-perf~\cite{DBLP:conf/middleware/WeichbrodtAK18}). 
STYX~\cite{DBLP:conf/cloud/WeiLLYG17} provides a trusted way for the content owner to provision the hardware cryptographic accelerator provided by the CPU of an untrusted cloud provider and thus benefit from Intel Quick Assist Technology (QAT~\cite{intel/whitepaper/qat-kpt}). 
Also, in STYX, an SGX enclave attested by the content owner is used to provision the TLS private key to the QAT engine, which is natively interfaced with OpenSSL~\cite{intel-qat-openssl}. 
This design suffers from the fact that interactions between the QAT engine and untrusted application are not limited to TLS~1.3 specific operations. 
As detailed in~\cite{DBLP:conf/eurosp/BhargavanBFOR17} w.r.t.\ Keyless SSL~\cite{DBLP:conf/trustcom/StebilaS15}, the use of such generic cryptographic operations may be exploited. 

Conclave~\cite{DBLP:conf/uss/HerwigGL20} takes a higher level approach by defining an architecture for securing a full service NGINX server, 
which runs on an untrusted infrastructure. 
Conclave presents two configurations for TLS~1.2 alone: 1) only the private key is protected by the TEE, or 2) the entire TLS (including the session keys) is protected by the TEE. 
In addition, just executing the TLS in a TEE as per Conclave is not viable both from  performance and operational perspectives.
Security-wise, Conclave uses Graphene which is a large library (more than 77000 LoC) and has a high probability for vulnerabilities as shown in~\cite{DBLP:conf/uss/CloostersRD20}. In contrast, \lurk\ has 3800 LoC and extends the private key protection to any authentication credentials used by TLS (including session resumption) without the need to deploy Graphene.
Also, unlike Conclave, \lurk{} provides anti-replay protection.

\subsection{TLS Protocol Extensions}

Similar to Keyless SSL, most previous works on TLS delegation~(e.g., see \cite{DBLP:conf/sp/LiangJDLWW14, 
DBLP:conf/trustcom/StebilaS15, DBLP:conf/eurosp/BhargavanBFOR17, DBLP:conf/cloud/WeiLLYG17}) are not designed for TLS~1.3 and suffer from the TLS~1.2 limitations~\cite{DBLP:conf/sp/LiangJDLWW14, DBLP:conf/trustcom/BoureanuMPAMFM20}.
Bhargavan et al.~\cite{DBLP:conf/eurosp/BhargavanBFOR17} provide delegation for Authenticated and Confidential Channel Establishment (ACCE) with TLS~1.3, yet there are two essential differences compared to our approach: ACCE is controlled by both ends (i.e., the client and the server), and it requires modifications to the TLS-record layer to achieve fine-grained access-rights for CDNs.
To the best of our knowledge, \lurk\ is the first design that provides a \emph{server-controlled} delegation \emph{specific to TLS~1.3}, without any modification to TLS~1.3, as well as leveraging TEEs for added trust.
Delegated credentials (DCs)~\cite{draft-ietf-tls-subcerts} is a TLS~1.3 extension  which eases the issuance of the authentication credential by a CDN provider.
However, the content owner delegates the authentication to the CDN, and the deployed credentials by the CDN remain exposed. 
In a DC deployment, \lurk{} can enable the CDN to protect the CDN authentication credentials (or the CDN can use any other TEE-based alternatives to protect the credentials). 
On the other hand, from the content owner perspective, \lurk{} makes DC unnecessary as the content owner's authentication credentials can be used without being shared to the CDN. 
This could be useful to ensure that existing/legacy TLS clients can authenticate the server; note that DC deployments require support/control from both the client and server sides (supported by the Firefox browser since 2019, and used by Cloudflare and Facebook services).

Boureanu et al.~\cite{DBLP:conf/trustcom/BoureanuMPAMFM20} used a similar design to \lurk\ but for TLS~1.2.
Most differences between Boureanu et al.~\cite{DBLP:conf/trustcom/BoureanuMPAMFM20} and us stem from TLS~1.3 being different from TLS~1.2. 
\lurk\ also offers several variants to interact with the \cs{} in the TEE to balance reasonable security vs.\ efficiency, which also addresses Boureanu et al.'s latency issues. 
In addition, \lurk\ leverages TEEs and provable security for further trust.

Various services  (referred to as middleboxes) provided by CDNs can only function when they have access to plaintext data, such as IDS, IPS, WAF, and L7 load balancing~\cite{de2020survey}.
In some previous proposals~\cite{DBLP:conf/uss/HerwigGL20, talos, brandao2021hardening}, these services cannot operate well within the CDN since they do not have access to plaintext data.
This problem was solved in TLS~1.2 by mcTLS~\cite{DBLP:conf/sigcomm/NaylorSVLBLPRS15}, but with significant overhead and  heavy modifications to TLS~1.2 handshake and record-layer.
It was also solved generically for any ACCE protocol, but again with significant overhead~\cite{DBLP:conf/eurosp/BhargavanBFOR17}.
However, \lurk\ solves this for TLS~1.3 without any modification to TLS~1.3 with an  acceptable level of overhead
(see Table~\ref{tab:relw_perf}). 

\section{Design Goals and Threat Model}
\label{sec:tm}
The main difference between \lurk{} and the standard TLS is that \lurk{} operates over 3 parties: \client{}, \engine{} and \cs. 
The \engine{} and \cs{} implement the server \server{}. \cs{} handles the authentication credentials and derives the necessary TLS secrets for \engine{} which interacts with the \client{}. 
Standard TLS instead operates over 2 parties: \client{} and \server{}.

\subsection{Goals}
The purpose of \lurk{} is to ensure security properties of a TLS communication between \client{} and \engine{}: 
providing authentication ensured by trustworthy credentials (private keys as well as PSK), and enabling PFS. 
In particular, \lurk{} ensures that a TLS communication between \client{} and \engine{} remains trustworthy even if \engine{} becomes compromised in the future as well as even if other \client{} or \engine{} on another edge server is compromised.   
These properties are also provided when \engine{} and \cs{} are operated  ``in situ'' or by a CDN.  To meet these properties the following goals are derived:
\begin{itemize}
	\item The \cs{} must provide read protection of the authentication credentials from a compromised \engine{} to prevent them from being used in future TLS sessions. 
With \lurk{}, a compromised \engine{} in the  ``in situ'' scenario or a network admin in the CDN scenario is not able to access the credentials via root access -- including dumping the memory. 
This differs from the standard TLS 1.3 threat model where \server{} must not be compromised, and anyone with privilege access to \server{} can access the credentials.
A direct consequence is that with \lurk{}, once an attacker has compromised \engine{}, and later when \engine{} recovers from this compromise, 
that attacker is not be able to interfere with any future TLS session -- including resumed sessions. 

\item The compromise of \engine{}, or \cs{} being administered by a CDN, does not provide any advantage to an attacker (although \cs{} is interfaced via LURK~\cite{draft-mglt-lurk-tls13}) compared to the use of a regular \client{}.
This is achieved by strongly binding the interactions between \engine{} and \cs{} to the TLS 1.3 exchange between \client{} and \engine{} via the freshness mechanism (see~\ref{fresh}) as well as enforcing \cs{} to operate over full TLS exchanges as opposed to hash of such exchanges.
This also provides a very efficient anti-replay protection for example when PFS is not properly enforced to meet a performance criterion and limit the generation of (EC)DHE keys (see \cite{RFC9325} Section 7.4).

\item 
\cs{} must be able to impose PFS by enforcing \cs{} to generate new and unique (EC)DHE keys for each TLS sessions.
This differs from the standard TLS 1.3 model, where in the case of a CDN, PFS is expected to be enforced by the CDN (we avoid this trust in CDNs). 

\end{itemize}

\subsection{Adversary Capabilities}

We assume \cs{} is trustworthy. 
In \lurk{}, \cs{} runs inside a TEE whose threat model assumes that TEE and its interfaces cannot be corrupted~\cite{DBLP:journals/iacr/CostanD16}.
\cs{} is expected to be developed with formal verification. The current 3800 LOC makes such assumptions realistic. 

The private key must be securely provisioned to \cs{}. 
This may involve the key being generated and distributed from a TEE~\cite{Blindfold} or the enclave being provisioned securely~\cite{intel-tls-ra}.
The latter is expected to be achieved with TLS 1.3 being implemented in the \cs{}.  
%
The attacker can control all 
\client{}s and \engine{}s, and interact either using TLS 1.3 or LURK. 
%
Note that, in a TLS session where the \client{} or \engine{} is under the control of the attacker, all session secrets are exposed to the attacker and the TLS session is trivially decrypted (but the private keys remain protected under \cs{}).

Regarding the network capabilities of the attacker, we also assume 
as per the  Dolev-Yao's threat model, that all public channels are accessible to the attacker to read, replay, block and inject messages.

\section{\lurk\ Design and Deployment Scenarios}
\label{sec:our-protocol}
In this section, we present our \lurk\ design, instantiated with TLS~1.3, including the protocol and example use cases for \engine{} and \cs{} based on their deployment. 

\subsection{\lurk\  -- Design}
\label{sec:design}
\begin{figure*}[htb]
 \centering
    \includegraphics[trim={0 35pt 0 0}, clip, width=0.85 \textwidth]{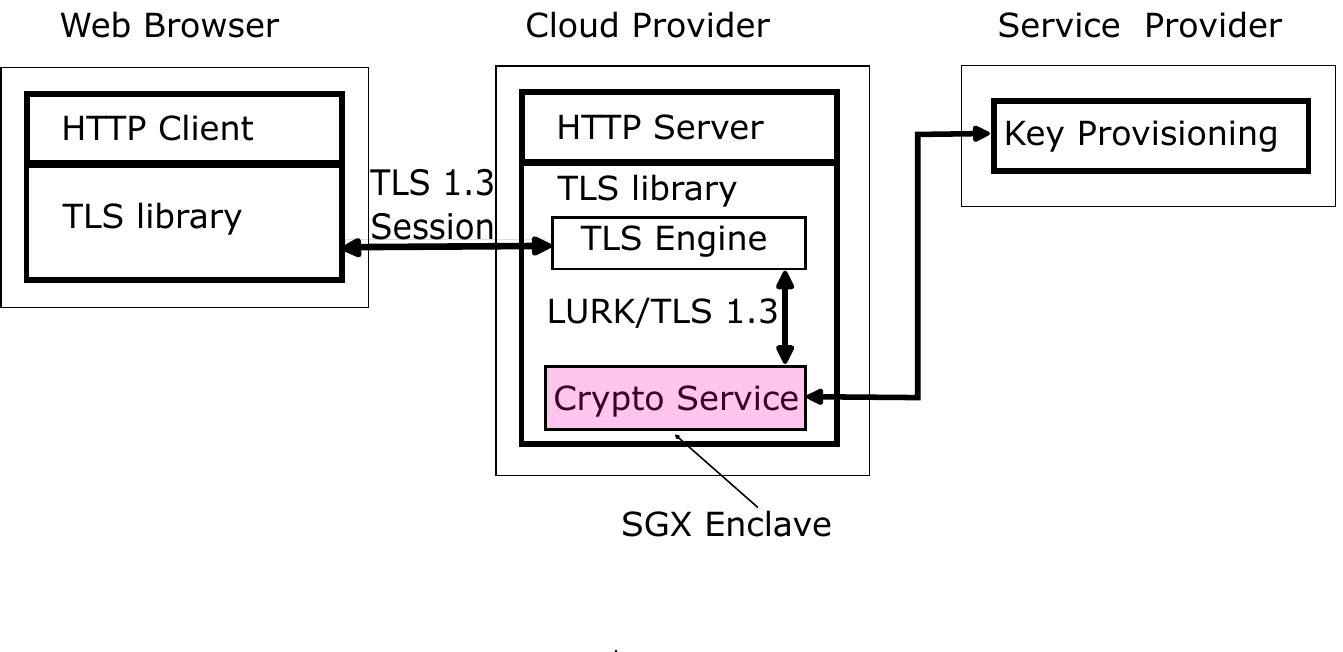}
    \caption{LURK-T entities: a TLS client \client{} (a regular web browser), a \lurk\  TLS Engine \engine{}, and a \lurk\ TLS Crypto Service \cs{} (both part of a third-party hosting provider), and a key provisioning server (under the content owner's control).}
\label{fig:lurk-overview}
\end{figure*}

\subhead{Components and the protocol}
\lurk{} involves the following entities: a TLS client \client{}, a \lurk\  TLS Engine \engine{}, and a \lurk\ TLS Crypto Service \cs{}; see Figure~\ref{fig:lurk-overview}.  
The last two are either collocated, or there is a pre-established, mutually authenticated and encrypted channel between them.
Such channel is expected to be implemented via a TLS library embedded into \cs{} that terminates into the TEE to prevent 
the communication between \engine{} and \cs{} being compromised by the node hosting \cs{}.
The key provisioning service is responsible for ensuring that the correct key is securely conveyed to  the \cs{} using a secure channel that terminates into the enclave. This can be achieved using solutions such as Blindfold~\cite{Blindfold}.

The purpose of TLS is to authenticate and agree on sessions keys so that \client{} and \engine{} can encrypt and exchange application data. 
The Key Schedule is responsible to generate the various secrets between \client{} and \server{} and includes, among others, the client/server handshake secrets ($h_{\client}$, $h_{\server}$) used to derive the keys that protect the TLS key exchange, the client/server application secrets ($a_{\client}$, $a_{\server}$) used to derive the keys that protect the application data, the session resumption secret ($r$) used to generate the PSK for authenticating \client{} and \server{}. 
The Key Schedule generates these secrets thanks to shared secrets such as PSK or (EC)DHE shared secret \ecdhesharesecret{} as well as the TLS handshake context \handshake.
The ClientHello.random \cnonce{} and ServerHello.random \snonce{} provide some randomness to generate these secrets.

TLS supports three basic key exchange modes: (EC)DHE (Diffie-Hellman over either finite fields or elliptic curves), PSK-only and PSK with (EC)DHE. 
The TLS (EC)DHE mode corresponds to  certificate-based authentication with \server{} being authenticated by \client{} by proving the ownership of a private key $\mathit{sk}$ via a signature over the TLS exchange context \handshake{} designated as $\ssign = \mathsf{Sign}_{\mathit{sk}} (\handshake)$. $\mathit{sk}$ constitutes the authentication credential.     
(EC)DHE ensures perfect forward secrecy, with \client{} and \server{} generating their respective private keys $u$ and $v$, exchanging their respective corresponding public key \ckeyshare{}=$g^u$ and \skeyshare{}=$g^v$ to generate a common (EC)DHE shared secret  \ecdhesharesecret{} = $g^{uv}$.  
The TLS PSK with (EC)DHE mode is based on a PSK shared between \client{} and \server{} as well as (EC)DHE, while the TLS PSK-only mode does not provide perfect forward secrecy.
In both cases, PSK is the authentication credential. 
The TLS (EC)DHE mode is commonly used on the web together with the TLS PSK in the (EC)DHE mode to resume TLS sessions.

\begin{figure*}
 \centering
  \begin{subfigure}[b]{\columnwidth}
    \centering
    \includegraphics[width=0.85 \textwidth]{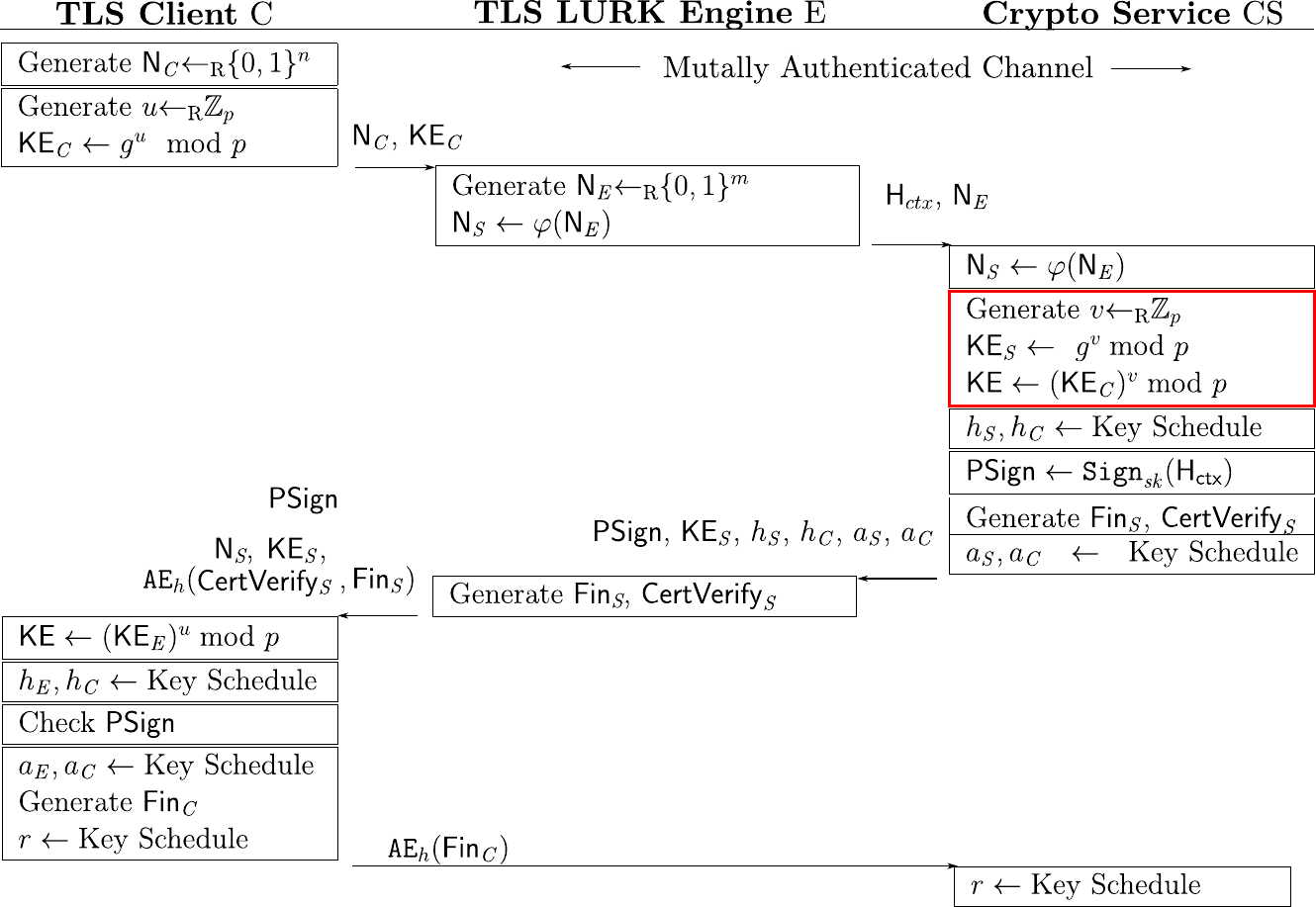}
    \caption{  \dhea, instantiated in  (EC)DHE mode }
    \label{fig:lurk-csdhea}
  \end{subfigure}
  \hfill
  \vspace*{0pt}
  \begin{subfigure}[b]{\columnwidth}
    \centering
    \includegraphics[width=0.85 \textwidth]{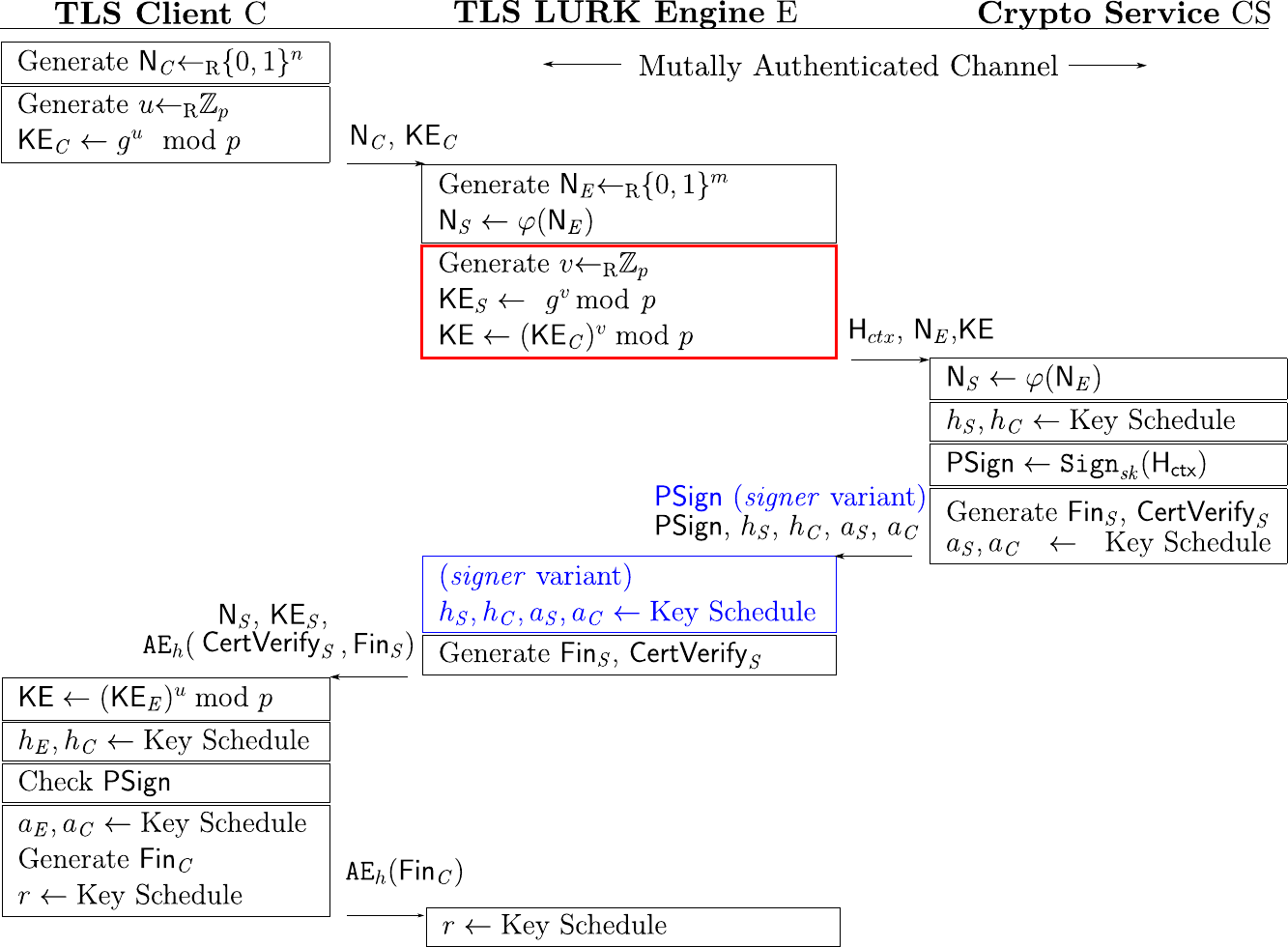}
    \caption{ \dhep, instantiated in  (EC)DHE mode}
    \label{fig:lurk-csdhep}
  \end{subfigure}
  \vspace*{-0.1cm}
\caption{The two variants of \lurk~  instantiated with TLS~1.3 in (EC)DHE Mode}
  \vspace*{-0.5cm}
\label{fig:lurk}
\end{figure*}

\lurk{} always assumes that the authentication credentials $\mathit{sk}$ or PSK are handled and hosted in \cs{}. As a result, in the TLS (EC)DHE mode, \ssign{} is generated by \cs{} and in the TLS PSK modes the Key Schedule is performed by \cs{}.   
On the other hand, \lurk{} enables the private (EC)DHE key to be generated either by \cs{} or \engine{} which leads to the respective variants \emph{\dhea} and \emph{\dhep} illustrated in  Figure~\ref{fig:lurk-csdhea}  and Figure~\ref{fig:lurk-csdhep}. The difference is highlighted in red.
\dhea{} provides higher (compared to \dhep{}) assurance on perfect forward secrecy (where TEE both protects the (EC)DHE key and attests the key is not reused), and on resumed TLS sessions (where TEE protects the PSK). 
The TLS execution between \client{} and  \cs, is actively proxied by \engine{}; i.e., \engine\ acts as the TLS server (\server{}) to \client, but  \engine\ does not have direct access to the private key of the origin (i.e., of  \cs) in order to generate the \ssign\   TLS message.

Note that Figure~\ref{fig:lurk-csdhea}  and Figure~\ref{fig:lurk-csdhep} depict \lurk{}  instantiated with TLS~1.3 in (EC)DHE mode. For PSK with (EC)DHE, the only difference stems from the key-derivation in TLS.  
We now describe in more detail the two main variants of  \lurk, each of these two modes. 
Description of the PSK-only mode is omitted as PSK-only can be easily derived from the PSK with (EC)DHE mode and this mode is rarely used in the web context, with even discussions at the IETF to deprecate that mode~\cite{rfc9257}.

\subhead{\dhea\  -- TLS (EC)DHE mode}
Following Figure~\ref{fig:lurk-csdhea}, \client{} initiates the TLS key exchange with \engine{} by sending a ClientHello message which contains the random \cnonce{} as well as the (EC)DHE public key. 

Upon receiving the ClientHello, \engine{} applies the freshness mechanism detailed in Section~\ref{fresh} to protect against replay and signing oracle attacks and provides the necessary handshake context to \cs{} to perform the Key Schedule and generate the signature.  
\engine{} generates a nonce \enonce{} and applies a pseudorandom function $\varphi$ to produce a bitstring denoted \snonce{}. In all variants and modes of \lurk, \enonce{} is deleted from memory at the end of the handshake. 
Then, \engine{} sends to \cs{} the whole of its view of the handshake $H_{ctx}$ (including \cnonce\ and \ckeyshare), the bitstring \enonce{}. 

\cs{} generates \snonce{} from \enonce{} similarly to \engine{}. 
As in \dhea, \cs{} generates the private (EC)DHE secret key $v$ and \skeyshare{}  and \ecdhesharesecret{}. \ecdhesharesecret{} is used together with \handshake{} by the Key Schedule to generate the handshake secrets ($h_{\client}$, $h_{\server}$). 
\cs{} generates the signature \ssign{}. 
\cs{} then generates the remaining handshake messages to update \handshake{} and have sufficient context to generate the application secrets ($a_{\client}$, $a_{\server}$). 
The formed messages are CertificateVerify (\certv), which contains the signature,  as well as the server Finished message (\sfin) which is a hash MAC of \handshake{}. 
Generating these message avoid an additional round trip between \engine{} and \cs{}.
\cs{} then provides \engine{} the signature \ssign{}, the (EC)DHE public key \skeyshare{} and handshake and application secrets.

Upon receiving \skeyshare{}, \engine{} generates and sends the ServerHello message with \skeyshare{} as well as the previously computed random \snonce{} to the TLS client \client{}.
\engine{} generates the CertificateVerify (\certv) and server Finished message (\sfin) and encrypts them with the session keys generated with the handshake secrets ($\algae_{h}$).  

\client{} performs its Key Scheduler (similarly to \server{}), checks the signature, generates the client Finished message (\cfin) and encrypts it ($\algae_{h}$) with session keys derived from the handshake secrets ($h_{\client}$, $h_{\server}$) before finally deriving the session resumption secret. 
%
Upon receiving \cfin{}, \engine{} forwards it to \cs{} so that \cs{} can generate the session resumption secret $r$  and the PSK, which will be used later during the session resumption.

\subhead{\dhep\ -- TLS (EC)DHE mode}
In this variant, \cs{} does not generate the (EC)DHE private key, which is instead generated by \engine{} (see Figure~\ref{fig:lurk-csdhep}).
\engine{}  generates \enonce{} exactly as in the \dhea\ variant. 
Then, \engine{} generates the (EC)DHE private key $v$ and associated public key \skeyshare{}, $g^v$, and computes \ecdhesharesecret{}, $g^{uv}$, which is provided to \cs{}, alongside the \handshake{} and \enonce{}. 
\cs{} then computes \snonce{} as in \dhea, 
initiates the Key Scheduler with \handshake{} and \ecdhesharesecret{} as inputs, computes \ssign{} and optionally the handshake and application secrets $h_{\server}$, $h_{\client}$, $a_{\server}$, $a_{\client}$ -- as these later secrets may also be generated by \engine{}.
These two sub-variants are represented in blue in Figure~\ref{fig:lurk-csdhep}, and are designated respectively as  ``\keyless{}'' or ``normal''.
The rest continues as in the \dhea\ variant.
Note that with  \dhep, when session resumption is enabled, the resumption secret $r$ is generated by \engine{} (not by \cs{}, and thus \cs{} is unable to guarantee its confidentiality). 

\subhead{\lurk\ variants with TLS PSK with (EC)DHE mode}
The main difference between the PSK with (EC)DHE mode and the (EC)DHE mode is that the former is used for session resumption. \lurk\ in PSK with (EC)DHE vs.\ (EC)DHE mode varies in as much as TLS~1.3 varies across these two modes. 
\lurk\ in PSK with (EC)DHE mode requires more exchanges between \engine{} and \cs{}. 
Typically, upon the reception of the ClientHello, \engine{} needs to check the PSK proposed by \client{} by performing a HMAC with a binder key derived from the PSK;  this binder key can be generated only by \cs{}, and \engine{} needs to request it from \cs{}. 
Once the PSK binders have been checked, \engine{} interacts with \cs{} to generate the various secrets as in (EC)DHE mode, but without \ssign{} being generated.

\subhead{Notes on \lurk's freshness function $\varphi$}
\label{fresh}
We derive the ``server-nonce'' \snonce{} by applying a non invertible PRF $\varphi$ instance to a nonce \enonce{} generated by \engine{} to prevent  replay attacks.
If an adversary $\adv$ collects plaintext information from a handshake, 
then $\adv$ will gather $N_C$, 
$\ckeyshare$  and  $N_S$ (from the channel in between $E$ and $C$). However, $\adv$ will not be able to  derive \enonce{} due to the non-invertible property of $\varphi$.
If later on, $\adv$ corrupts \engine{}, $\adv$ will not find the old $N_E$ nonce in 
$\engine$'s 
memory; we require that $N_E$ be deleted from $E$'s memory at the end of its use. Exhaustive search of the right $N_E$ would also be exponential in the size of the domain of $\phi$, so it will be impossible for our polynomial attackers, and thus preventing replay attacks as \enonce{} is necessary for the exchange.

\subsection{\lurk\ - Use Cases and Deployment Scenarios}
\label{sec:uc}
We consider different  deployment scenarios for \lurk{} as discussed below. 
The management of TLS is impacted by the management of TEE (with attestation) as well as the management of the long term private key; other aspects of TLS are not impacted. 
The CS Manager is the entity responsible to administrate and provision \cs{}. 
Unless the private keys are generated inside the enclave, the CS is responsible to provision the \cs{} with the secret key. 
Securely provisioning the enclave can be achieved by combining attestation and terminating the communication within the enclave. 
The enclave implementation must be verified by the CS manager  (requiring \cs{} code to be open-sourced). 
It also likely requires a TLS library being embedded into the enclave.  
Similarly, as only the TLS library is impacted, \lurk{} enables the CDN to continue providing added services, and as such, keeps TLS a multi-party TLS.  

\begin{figure}[hbt]
  \centering
  \begin{subfigure}[b]{0.45\columnwidth}
    \centering
    \includegraphics[width=\textwidth]{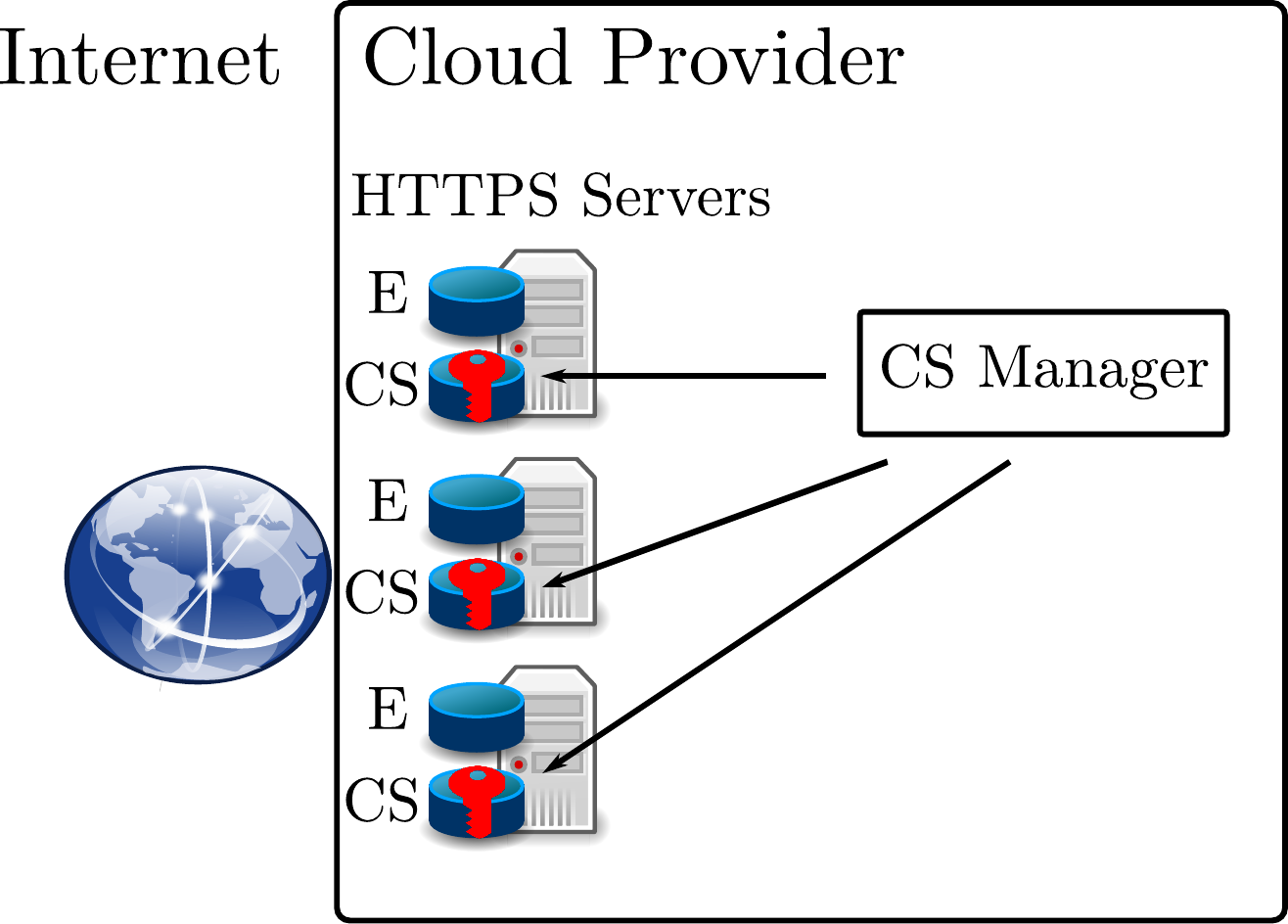}
    \caption{Distributed \cs{}}
    \label{fig:multi-cs}
  \end{subfigure}
  \hfill
  \begin{subfigure}[b]{0.45\columnwidth}
    \centering
    \includegraphics[width=\textwidth]{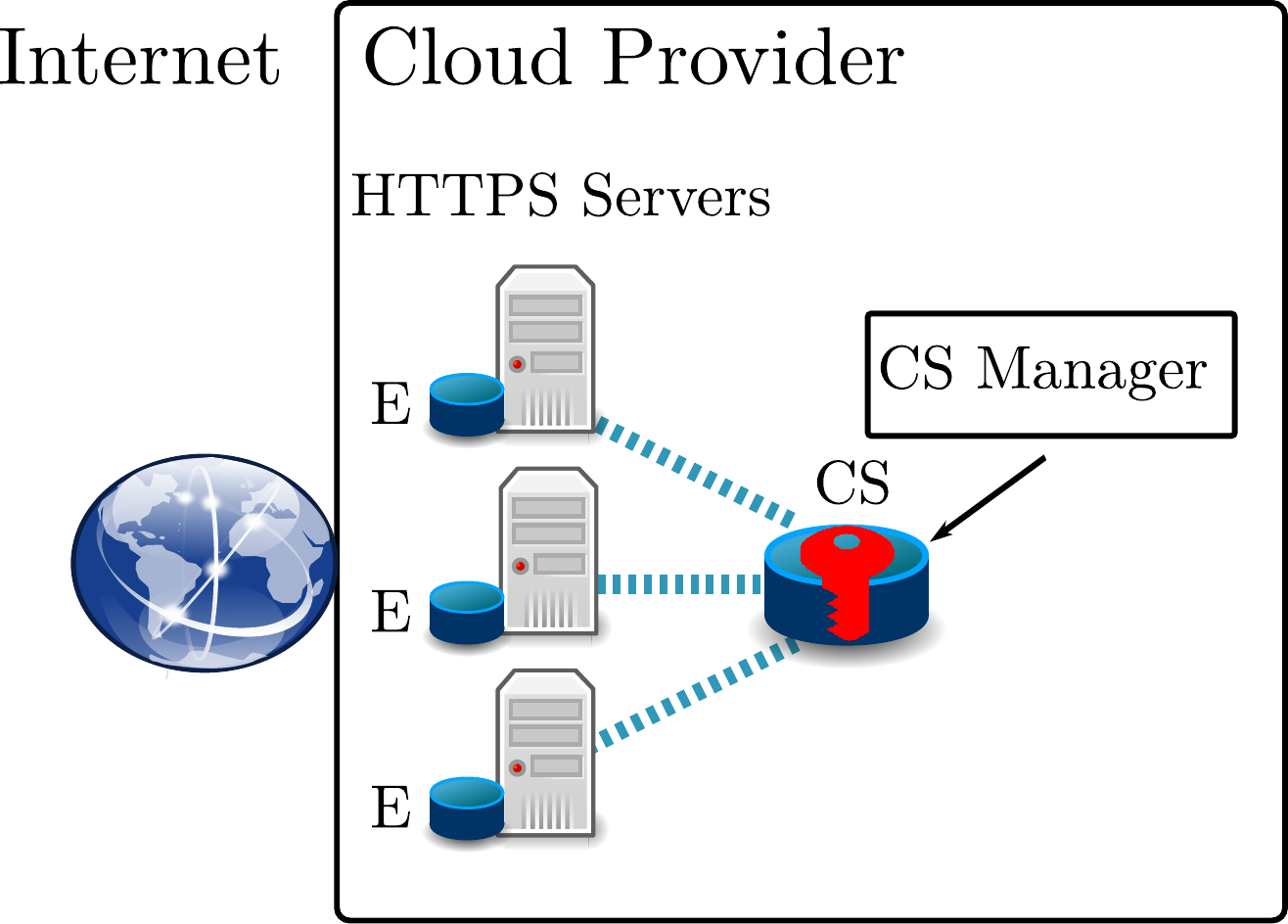}
    \caption{Centralized \cs{}}
    \label{fig:centralized-cs}
  \end{subfigure}
  \hfill
  \begin{subfigure}[b]{0.45\columnwidth}
    \centering
    \includegraphics[width=\textwidth]{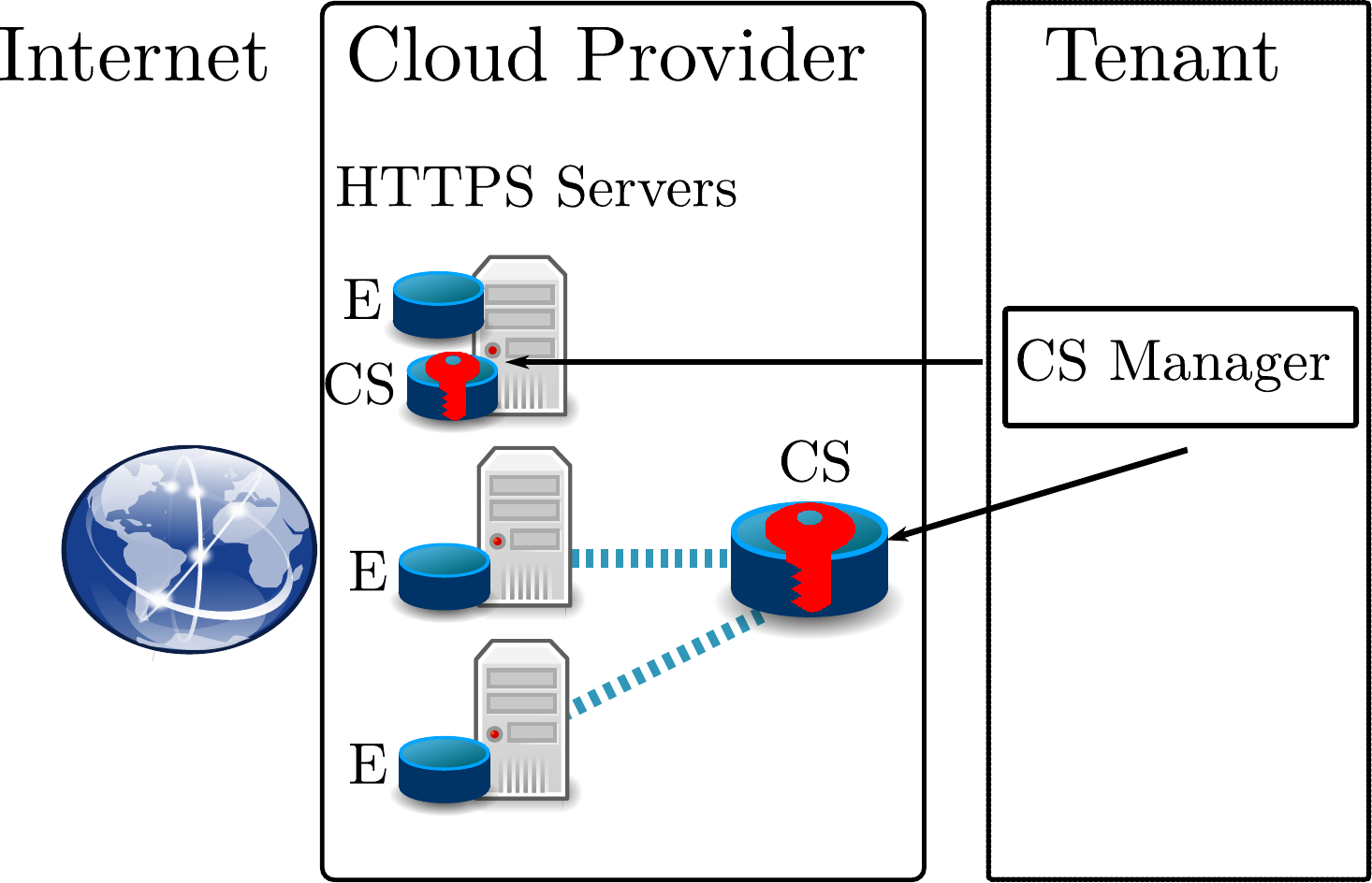}
    \caption{Tenant-controlled \cs{}}
    \label{fig:tenant-cs}
  \end{subfigure}
  \vspace*{-0.2cm}
  \caption{CS deployment use cases}
    \vspace*{-0.6cm}
  \label{fig:use-cases}
\end{figure}

\subhead{Deployments driven by CDN providers}
Figure~\ref{fig:multi-cs} shows the case where \lurk{} is deployed as a substitute of TLS~1.3 libraries. In this  case, the server-side TLS libraries are replaced both by 
\engine{} and \cs{}.
The main challenge associated with this case is that 
CDN providers will need to manage (and provision) multiple instances of \cs{}. 
Figure~\ref{fig:centralized-cs} shows the case of a more centralized infrastructure, with just one  \cs{} with an SGX enclave 
communicating securely
with multiple \engine{}s. In both cases,  it remains crucial to implement an attestation-ready provision of the \cs{}. As the attestation is to be performed by the CDN provider within its own network, DCAP
seems appropriate in combination with TLS-RA~\cite{intel-tls-ra}. 

\subhead{Deployments driven by content owners}
Figure~\ref{fig:tenant-cs} shows the case where \cs{} is provisioned by a CDN tenant, such as a content owner. 
Therein, \cs{} is likely to be implemented by a third party (\cs{} developer), trusted by the content owner and  the cloud provider -- e.g., with open source code. 
The tenant will need to perform an attestation of the \cs{}, e.g., using Intel IAS~\cite{intel-sgx-epid-attestation
}; this should use a group signature in order for the tenant not to find out the identifier of the exact CPU running the \cs{}. This is also likely to be combined with RA-TLS~\cite{intel-tls-ra}. 

Finally, note that from a tenant's perspective there is a little difference between instantiating a centralized \cs{}, or multiple \cs{} instances; the difference is mostly in the way \cs{} is implemented, which can also be checked by the tenant via attestation. 

\section{System Implementation}
\label{sec:impl}
\label{sec:implementation}
In this section, we describe our implementation\footnote{Available
from \url{https://github.com/lurk-t/}}  of \cs{} and \engine{} based on OpenSSL. 
\cs{} centralizes the cryptographic operations. However, OpenSSL has not been developed with such a centralized cryptographic architecture and instead performs TLS operations sequentially.
Thus, following the OpenSSL design would lead to numerous interactions between \engine{} and \cs, and degrade performance, especially when interactions are between the Rich Execution Environment (REE) and TEE.
In particular, for SGX enclaves, the interaction between TEE and REE results in 8,200 - 17,000 cycles overhead, as opposed to 150 cycles for a standard system call~\cite{DBLP:conf/isca/WeisseBA17}.
We also balance the compliance to the specification of the LURK extension for TLS~1.3~\cite{draft-mglt-lurk-tls13} and changes to OpenSSL to ease the maintainability of our code. 
As a result, we implement \engine{} by updating 184 lines of the OpenSSL code and introducing a maximum of 2 additional ECALLs compared to the LURK specification~\cite{draft-mglt-lurk-tls13}. 
Our \cs{} implementation contains 33 files with 3867 LoC.
\subsection{Crypto Service (\cs{})}
\label{sec:impl:cs}
We implemented \cs{} in an SGX enclave based on Intel SDK version 2.13. 
We had several options regarding the cryptographic library. 
While some cryptographic libraries 
support terminating the TLS connection inside SGX, we did not use them since they are either not maintained~\cite{talos}, or not fully compatible with OpenSSL.\footnote{\url{https://www.wolfssl.com/wolfssl-with-intel-sgx}}
We chose the actively maintained Intel SGX-SSL~\cite{SGX-SSL} that compiles OpenSSL source code as-is to create SGX compatible APIs (ensuring compatibility and easy upgrades with future versions of OpenSSL). 
However, SGX-SSL has limited functionalities. 
For example, it does not support terminating TLS inside SGX and lacks all the TLS and network related structures. 
Therefore, part of the \cs{} implementation mimics the TLS specific functions implemented by OpenSSL using lower-level APIs and structures supported by SGX-SSL
(we use OpenSSL 1.1.1g for SGX-SSL). 

\subsubsection{\cs{} in TLS (EC)DHE mode}
\label{sec:impl:ecdhe}
\begin{figure}[htbp]
\centerline{\includegraphics [width=0.8\columnwidth] {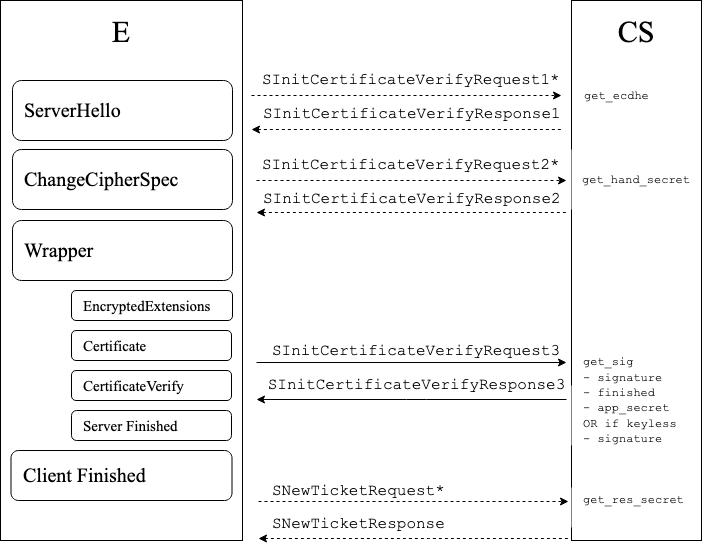}}
\caption{Messages between \engine{} and \cs{} for the (EC)DHE mode. * designates an optional exchange depending on the \cs{} configuration}
\label{fig:exchange_order}
\end{figure}

\cs{} is responsible for generating the different parts of the handshake such as the signature, and optionally \textemdash\ depending whether \cs{} operates in the \dhea{} or \dhep{} variant \textemdash\  (EC)DHE keys and secrets as detailed in Section~\ref{sec:design}. 
Our implementation supports all these variants as depicted in Figure~\ref{fig:exchange_order} which details the exchanges between \engine{} and \cs{}.
While our design defines a single SInitCertificateVerify exchange~\cite{draft-mglt-lurk-tls13}, our implementation, when necessary (depending on the \cs{} configuration), repeats up to 3 times that exchange in order to retrieve different pieces of information (depending on the \cs{} configuration, see Figure~\ref{fig:exchange_order}). 
Table~\ref{tab:cs-conf} shows the supported \cs{} configurations, and for each one, which entity (\engine{} or \cs{}) generates the (EC)DHE or the secrets $h$, $a$ and  $r$.
Binder keys and signature are always generated by \cs{} in their respective (EC)DHE or PSK with (EC)DHE modes. 

\begin{table}[htb]
\centering
\begin{tabular}{|l|c|c|c|c|}
\hline
\cs{} config (Cert) & $CS_{cert}^{dhe, r }$ & $CS_{cert}^{dhe}$ & $CS_{cert}$ & $CS_{cert}^{\keyless{}}$\\
\hline
(EC)DHE       & \cs{}                 &  \cs{}            & \engine     & \engine  \\
handshake     & \cs{}                 &  \cs{}            & \cs{}       & \engine \\
application   & \cs{}                 &  \cs{}            &   \cs{}     & \engine \\
resumption    & \cs{}                 &   --              & \cs{}       & \engine \\
\#ECALLs      &   4                   &  3                &  2          & 1       \\ 
\hline
\cs{} config (PSK) & $CS_{psk}^{dhe, r}$ & $CS_{psk}^{dhe}$ & $CS_{psk}^{r}$ & $CS_{psk}$ \\
\hline
(EC)DHE       & \cs                 & \cs              & \engine        & \engine \\
handshake     & \cs                 & \cs              & \cs            &  \cs    \\
application   & \cs                 & \cs              & \cs            &  \cs    \\
resumption    & \cs                 &  --              & \cs            &  --     \\
\#ECALLs      & 5                   & 4                &  4             &  3      \\
\hline
\end{tabular}
\caption{\cs{} configurations indicating where (EC)DHE or secrets are generated (when generated) and  associated number of ECALLs by our implementation} 
\label{tab:cs-conf}
\end{table}

\subhead{Generating (EC)DHE}
In the \dhea{} variant, \cs{} generates the (EC)DHE private key for \server{}. 
\engine{} retrieves \server's (EC)DHE public key with an additional SInitCertificateVerify1 exchange (see Figure~\ref{fig:exchange_order}). 
\cs{} generates the (EC)DHE shared secret using \client{}'s (EC)DHE public key and the  \server's (EC)DHE private key -- that is kept secret by \cs{}. 
This is implemented with our \texttt{get\_ecdhe} function which represents an additional ECALL compared to the LURK specification. 

\subhead{Generating $h$ and $a$}
When \cs{} is configured to generate $h$, \engine{} performs an additional SInitCertificateVerify2 exchange to retrieve handshake secrets $h_{\client}$ and $h_{\server}$  (see Figure~\ref{fig:exchange_order}): 
our function \texttt{get\_hand\_secret} takes the ClientHello to ServerHello messages as inputs and returns $h$.  
This represents an additional ECALL compared to the LURK specification.
When \cs{} is configured to generate $a$, both $a_{\client}$ and $a_{\server}$ are generated together with the signature in our \texttt{get\_sig} function (SInitCertificateVerify3).
\texttt{get\_sig} takes the ClientHello to EncryptedExtension messages, generates the signature, completes the TLS handshake by generating the CertificateVerify and the server Finished messages to compute $a$. 
In contrast, in the \emph{\keyless{}} configuration, \texttt{get\_sig} only generates the signature; therefore, in this case, our implementation fully matches the LURK specification with a single ECALL.

\subhead{Session resumption}
When session resumption is enabled, a new session ticket  is retrieved via a SNewTicket exchange. 
This exchange provides the full TLS handshake (from ClientHello to client Finished) and a nonce to the \cs. 
Our implementation generates a stateful ticket in which \cs{} generates the resumption master secret $r$ and, subsequently, uses it for generating the PSK. 
\cs{} stores the PSK and the \lurk{} session ID (that is used as a PSK ID) in the TEE. 
Therefore, \engine{} caches the \lurk{} session ID as a PSK ID to further identify the PSK. 
OpenSSL handles the generation of the NewSessionTickets messages as well as the ability to bind a ticket in a resumed session to the PSK generated in a previous TLS session. 
To fully reuse OpenSSL ticket management functions, the PSK ID is stored where OpenSSL used to store the clear text PSK. 

\subsubsection{\cs{} in TLS PSK with (EC)DHE mode}

During a session resumption, our implementation blocks OpenSSL from accessing the PSK, and instead \engine{} sends a SInitEarlySecretRequest to \cs{}.
This exchange provides the PSK ID so \cs{} can restore the PSK and initiate a Key Schedule and return the binder key.
Similar to Section~\ref{sec:impl:ecdhe}, the specified SHanshakeAndApp 
is implemented in 3 ECALLs when \cs{} generates the (EC)DHE (\texttt{get\_\-ecdhe}, \texttt{get\_hand\_secret} and \texttt{get\_\-app\_\-secret}), or 2 ECALLs when \engine{} generates the (EC)DHE (\texttt{get\_\-hand\_\-secret} and \texttt{get\_\-app\_\-secret}).
Generation of the resumption secrets $r$ by \cs{} requires an additional ECALL (\texttt{get\_\-res\_\-secret}).
\subsection{ TLS Engine (\engine{})}
\label{sec:impl:openssl}
\engine{}, which is based on OpenSSL~1.1.1g, is implemented by updating 9 C files out of the 44 files in the SSL directory.
Upon configuration, \engine{} executes the native OpenSSL function or initiates an exchange with \cs{}. 
OpenSSL defines two core structures: \texttt{SSL} and \texttt{SSL\_CTX}. 
\texttt{SSL} is created for each new TLS connection and contains all TLS sessions' context (e.g., cipher suite, session, secrets, etc). 
The communication between \engine{} and \cs{}, is handled via the \texttt{LURKRequest} and the \texttt{LURKResponse} structures added to the \texttt{SSL}.

\texttt{SSL\_CTX} contains the information common to all \texttt{SSL} structures (e.g., session resumption and the number of new TLS connections).
Typically, \client{} and \server{} create one \texttt{SSL\_CTX} structure and reuse it for all their TLS connections.
Since \cs{} is shared across all TLS connections, it is instantiated at the creation of \texttt{SSL\_CTX}.
Thus, initiating the enclave -- which is a time-consuming -- only happens once for \server{}.

To apply the freshness function, we need both the full TLS messages as well as the ServerHello.random \enonce{} (before applying the freshness function) -- see Section~\ref{fresh}.
However, by default, OpenSSL prevents the access to the TLS messages as it continuously hashes the  TLS messages to avoid storing large handshake data. 
To overcome this, our implementation stores the value of ServerHello.random (\enonce{} generated by OpenSSL) as well as handshake data. 
When OpenSSL generates \enonce{}, it is intercepted by the freshness function, stored in the LURKRequest, and replaced by \snonce{} so OpenSSL proceeds to the generation of the ServerHello using \snonce{}. Later on, \cs{} checks $\snonce{} = \varphi( \enonce{} )$, with \snonce{} being the ServerHello.random (\snonce{}) in the TLS message and \enonce{} the stored value.

Finally, \cs{} is integrated into \engine{} as an external library. 
We successfully linked (by updating OpenSSL Makefile) and tested our library for dynamic and static versions of OpenSSL.

\section{Performance Evaluation}
\label{sec:perf}
\subsection{Methodology for Measuring \lurk{} TLS Overhead over OpenSSL}
\label{sec:TLS-Evaluation}
In this section, we report the performance overhead of our TLS library.
The performance is measured in terms of TLS Key EXchange per second (KEX/s), following the methodology used in RUST TLS performance evaluation.\footnote{\url{https://jbp.io/2019/07/02/rustls-vs-openssl-handshake-performance.html}}
$\Delta_{KEX}$ expresses the relative difference in terms of KEX/s between \lurk{} TLS and the native OpenSSL TLS. 
In particular, $\Delta_{KEX}$ is expressed as a percentage for a given configuration $conf$ which represents the TLS cipher suites (see Table~\ref{tab:tls-conf}) and the tasks performed by \cs{} (see Table~\ref{tab:cs-conf}).
\begin{center}
$\Delta_{KEX} = \frac{| KEX_{\lurk} - KEX_{OpenSSL}|_{conf}}{KEX_{OpenSSL}}$
\end{center}
Our measurements are performed on an Intel i9-9900K CPU @3.60GHz over Ubuntu 18.04 LTS and we took the average time after performing 
10,000 handshakes.

\begin{table}[htb!]
\centering
\begin{tabular}{|l|l||l |}
\hline
Notation       & Description        & KEX/s\\ 
\hline
RSA-2048 & (prime256v1, RSA-2048)   & 1715 \\
RSA-3072 & (prime256v1, RSA-3072)   & 316 \\
RSA-4096 & (secp384, RSA-4096)      & 243 \\
P-256    & (prime256v1, P-256)      & 5251 \\
P-384    & (secp384, P-384)         & 496 \\
Ed25519      & (X25519, Ed25519)    & 6113  \\
Ed448        & (X448, Ed448)        & 1251 \\
\hline
\end{tabular}
\caption{Native OpenSSL TLS key exchange (KEX) performance for different cipher suites}
\label{tab:tls-conf}
\end{table}

\subhead{TLS cipher suites configuration}
We base our selection of cipher suites in Table~\ref{tab:tls-conf}, on Mozilla's \emph{modern compatibility} configuration which recommends ECDSA (P-256) or RSA-2048 combined with X25519, prime256v1, secp384r1.We added ECDSA (P-384) and Ed25519, projecting the measurement toward long-term deployments.

\subhead{CS configurations}
Besides TLS cipher suites, we measured various configurations for \cs{}. 
The primary purpose of \cs{} is to protect authentication credentials (private key $cert$ or $psk$). 
In the (EC)DHE mode (expressed as $cert$), session resumption may be enabled (expressed as $r$), so future handshakes may use the PSK with (EC)DHE mode.
To remain coherent across sessions in terms of PFS, we only considered the PSK with (EC)DHE mode (expressed as $psk$).
As mentioned in Section~\ref{sec:design}, the PSK is derived from the generated (EC)DHE shared secret. 
Thus, the PSK used for the session resumption can only remain confidential in a \dhea{} variant (e.g., the \cs{} generates the (EC)DHE private key). This is expressed with the following configuration $CS_{cert}^{dhe, r}$.
Of course, without session resumption, \dhea{} or \dhep{} variants are valid configuration 
expressed as $CS_{cert}^{dhe}$, $CS_{cert}$ or $CS_{cert}^{\keyless{}}$ (when only the signature \ssign{} is generated, see Section~\ref{sec:design}).  
In the PSK with (EC)DHE mode, and unlike the (EC)DHE mode, session resumption may be enabled with both \dhea{} or \dhep{} variants, and 
Table~\ref{tab:cs-conf} summarizes the meaningful \cs{} configurations with the associated number of ECALLs.

\subsection{Experimental Measurements of \lurk{} TLS Overhead over OpenSSL}
\label{sec:TLS-measure}
\begin{figure}
   \centering
    \includegraphics[width=0.6\columnwidth]{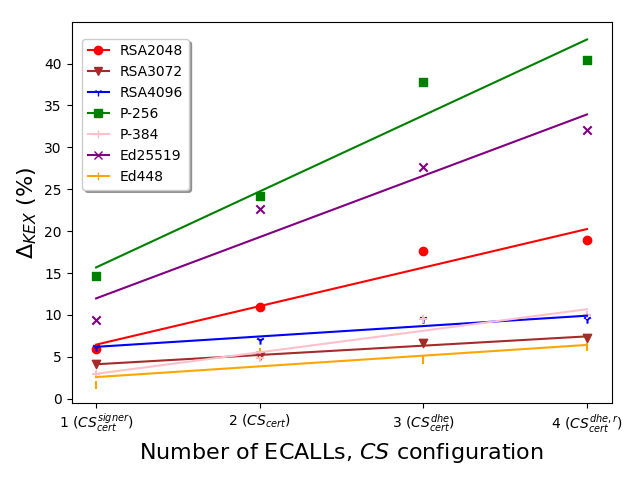}
    \caption{KEX \lurk{} TLS relative overhead over OpenSSL ($\Delta_{KEX}$) in (EC)DHE mode. Measured values are linked using a linear regression.}
    \label{fig:perf-ecdhe}
\end{figure}

\subhead{(EC)DHE mode} Figure~\ref{fig:perf-ecdhe} depicts $\Delta_{KEX}$ as a function of the number of ECALLs which characterizes \cs{} configuration (see Table~\ref{tab:cs-conf}). 
As shown in Figure~\ref{fig:perf-ecdhe}, ECALLs do not equally affect all cipher suites and $\Delta_{KEX}$ does not linearly increase with the number of ECALLs. 
However, as per Table~\ref{tab:tls-conf}, cipher suites that require more resources (RSA-3072, RSA-4096, P-384, Ed448), seem less impacted by \lurk{} TLS and their overhead depends more linearly on the number of ECALLs. 
A possible explanation is a low ratio of allocated slots by the scheduler which results in either an interruption or an exitless process wasting the remaining allocated cycles.  
With our current configurations, the measured overhead for Ed448, RSA-3072, P-384 and RSA-4096 is low (between 1.2\% and 10\%) and the number of ECALLs have very little impact. 
Other cipher suites (including RSA-2048) are more impacted by the number of ECALLs. 
Nonetheless, our implementation presents a higher overhead for the P-256 and Ed25519 cipher suites.
P-256 has up to 39.7\% overhead when (EC)DHE is performed by \cs{} and 14.7\% in the \emph{\keyless{}} configuration. 
Ed25519 is less affected in the \dhep{} variant (less than 23\%) compared to the \dhea{} variant (up to 33\%).
Finally, the \emph{\keyless{}} configuration provides an apparently acceptable overhead (17\% for P-384, 7.6\% for RSA-2048, less than 4.3\% for the others). 

\begin{figure}
\centering
    \includegraphics[width=0.6\columnwidth, height=6cm]{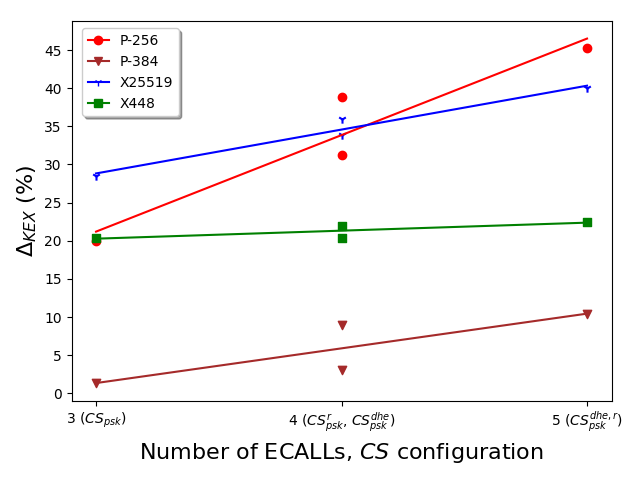}
    \caption{KEX \lurk{} TLS relative overhead over OpenSSL ($\Delta_{KEX}$) in PSK with (EC)DHE mode.}
    \label{fig:perf-psk}
\end{figure}

\subhead{PSK with (EC)DHE mode} Similar to the (EC)DHE mode, Figure~\ref{fig:perf-psk} shows the most efficient ciphers (P-256, X25519) are more impacted by the number of ECALLs than the others -- such as P-384 and X448. 
Overall, the preliminary measurements of our implementation show encouraging results with a limited and acceptable overhead.

The observed overhead might be further improved (both for (EC)DHE and PSK with (EC)DHE modes).
Firstly, we can reduce the number of ECALLs, which may incur major modifications to the OpenSSL architecture, and may affect the case of Encrypted Extension (see Section~\ref{sec:implementation}). 
Secondly, we can aggregate multiple \lurk\ requests in each ECALL. 
The optimal number of \lurk\ requests that need to be aggregated is expected to depend on the CPU, the cipher suite, and the \cs{} configuration. 
The optimum performance will be reached when multiple operations can be completed within the allocated number of cycles, minimizing the number of unused cycles. This is likely to benefit Ed25519 or P-256.

\subsection{SGX Vulnerabilities Mitigation Overhead}
\label{sec:perf-cs-mitigation}

In this section, we discuss the overhead associated to the available mitigations -- micro code or SDK~\cite{intel-affected-cpu} -- of SGX vulnerabilities for our CPU. 
The discussion in Section~\ref{sec:TLS-measure} considers the default SGX configuration; that is without vulnerabilities.  
While we expect future CPUs to address the currently known vulnerabilities --leading to the performances of Section~\ref{sec:TLS-measure}-- we also anticipate new vulnerabilities to be disclosed and their mitigation will come with an additional overhead for the cloud provider.
Note that previous proposals did not measure performance with these added security measures (which incur significant performance overhead).

According to Intel~\cite{intel-affected-cpu}, our CPU 
remains vulnerable to Special Register Buffer Data Sampling (SRBDS)~\cite{kernel-srdbs} and CrossTalk attack~\cite{crosstalk} for which Intel provides a microcode update.
Similarly, our CPU is vulnerable to Load Value Injection (LVI)~\cite{DBLP:conf/sp/BulckM0LMGYSGP20} which we respectively mitigate both via the SDK or via the SGX-SSL \texttt{cve\_2020\_0551\_load} ($ld$) or \texttt{cve\_2020\_0551\_cf} ($cf$)~\cite{SGX-SSL}.

Similar to Section~\ref{sec:TLS-Evaluation}, the performance is measured in terms of the number of KEX/s. 
In our case, the overhead of the microcodes -- SRBDS -- is negligible while the one of the  SDK and SGX-SSL -- for LVI -- is not. 
Table~\ref{tab:perf-cs-ecdhe} summarizes our measurements for each cipher suite. 
From the table, it is clear that for a given SGX configuration, the overhead increases with the number of operations performed by \cs{}. However, for a given cipher suite, we could not correlate the number of operations to the expected overhead. 

\begin{table}
\centering
\begin{tabular}{|l|c|c|c|}
\hline
          & $SGX^{SRDBS, cf}$ & $SGX^{SRDBS, ld}$ & $SGX^{default}$ \\
\hline                                           
RSA-2048  & 1162 & 132 & 1715 \\
RSA-3072  & 282  & 35 & 316\\
RSA-4096  & 181 & 17 & 243\\
P-256     & 2353 & 971   & 5251\\
P-384     & 277 & 48   & 496\\
Ed25519   & 3312 & 909 & 6113\\
Ed448     & 1081  & 78 & 1251\\
\hline
\end{tabular}
\caption{SGX performances (KEX/s) with SRDBS and LVI mitigation enabled versus default SGX for \cs{} configured with  $CS_{cert}^{dhe, r }$.}
\label{tab:perf-cs-ecdhe}
  
\end{table}

\subsection{\lurk{} TLS Overhead for HTTPS}
\label{sec:perf-https}
We developed a multithreaded HTTPS server in RUST (using OpenSSL). Subsequently, we modified it to use \lurk{} TLS to confirm that migration to \lurk{} TLS is easy and can be used in other programming languages that support OpenSSL (in this case RUST). 
Similar to $\Delta_{KEX}$ in Section~\ref{sec:TLS-Evaluation}, we measure  $\Delta_{HTTPS}$ (see Table~\ref{tab:cs-https-lan}), the overhead of \lurk{} TLS over HTTPS with OpenSSL by measuring the relative difference in requests by second of various file sizes being served. 
	To do so, we modified the benchmark tool wrk\footnote{\url{https://github.com/wg/wrk}}
to  force select TLS~1.3 as well as to be able to specify a specific cipher suite. 
The HTTPS server and benchmark tools are published as open source.\footnote{\url{https://github.com/lurk-t/https}}

We measure the number of HTTPS requests per second performed by wrk with 10 parallel connections to introduce some concurrency similarly to~\cite{DBLP:conf/cloud/WeiLLYG17} -- though in the measurements, we did not observe a significant difference between 10 and a single connection.
To prevent underestimating the impact of \lurk{} TLS, we considered our LAN with a 10 ms latency 
with 100 MB bandwidth that reflects the interactions with a NIC while lowering the impact of the latency.
Similarly, the download file is always cached in the memory of the HTTPS server, and thus, reducing \server's latency (by avoiding reading from the hard drive). 
We limit \cs's configuration to the most secure configuration which has the highest overhead ($CS_{cert}^{dhe, r }$). 
Moreover, to consider the SGX vulnerabilities, we performed the same measurements on the fully mitigation-enabled SGX (enabling $ld$ and $SRDBS$), which has the most overhead.

The measurements in Table~\ref{tab:cs-https-lan} show that even with $CS_{cert}^{dhe, r }$, the overhead is always negligible when downloading 1MB (or larger)  files. 
In other words, for such  files, the transfers overtake the overhead introduced by the \lurk{} TLS handshake. 
For file sizes lower than 100KB, \lurk{} TLS seems to introduce a slight overhead, but we are not able to find a clear relation with the file size, and  the overhead seems primarily determined by the cipher suite. 
Similar to the measurements of \cs{} in Section~\ref{sec:TLS-measure}, P-384 and Ed448 seem to provide better performances compared to other cipher suites.
Our reported measurements are valid from both the client and server perspectives. 
Note that, resource wise, \server{}'s \lurk{} overhead is the one reported in Section~\ref{sec:TLS-measure}.

When the mitigations ($ld$ and $SRDBS$) are enabled, the measurements show that P-256, Ed25519, RSA-2048 provide a negligible overhead for 1MB files. 
For other cipher suites, such a pivot seems to occur for file sizes over 1 MB. 

\begin{table}
\centering
\begin{subtable}[h]{\columnwidth}\centering
  \begin{tabular}{|c|c|c|c|c|c|}
\hline
$\Delta_{HTTPS}$ &  0KB  & 1KB  & 10KB & 100KB & 1MB   \\   
\hline                                                       
RSA-2048           & 16.0  & 17.7 & 8.9	  & 7.9   & 0\\ 
RSA-4096           & 7.4   & 8.7  & 7.2   &	8.6   & 0\\ 
P-256              & 5.0   & 4.0  & 4.2   &	10.8  & 0\\ 
P-384              & 5.7   & 8.1  & 0.8   & -3.0  & 0\\ 
Ed25519            & 10.9  & 13.9 & 3.4   &  3.4  & 0.1\\ 
Ed448              & 0     & 0    & 0     &  2.9  & 0 \\ 
\hline                                                       
  \end{tabular}
  \caption{Default SGX: \lurk{} TLS overhead in term of HTTP request/s is negligible for files larger than 1MB. }
  \label{tab:cs-https-sgx}
\end{subtable}
\begin{subtable}[h]{\columnwidth}\centering
\begin{tabular}{|c|c|c|c|c|c|}
\hline
$\Delta_{HTTPS}$ & 0KB & 1KB    & 10KB & 100KB & 1MB   \\   
\hline                                                       
RSA-2048 & 88.1 &  88.0 &  86.3 &  83.7 &  0.6 \\ 
RSA-3072 & 86.6 &  86.7 &  86.5 &  86.5 &  69.1 \\ 
RSA-4096 & 90.5 &  90.7 &  90.6 &  90.8 &  85.2 \\ 
P-256    & 76.6 &  76.4 &  78.2 &  78.8 &  41.7 \\ 
P-384    & 78.3 &  78.9 &  79.5 &  79.8 &  60.4 \\ 
Ed25519  & 63.2 &  63.4 &  58.1 &  38.5 &  0 \\ 
Ed448    & 78.1 &  77.9 &  78.5 &  82.7 &  38.2 \\ 
\hline                                                       
  \end{tabular}
  \caption{Mitigation-enbaled SGX ($ld$ and $S$)}
  \label{tab:cs-https-lvi}
\end{subtable} 
\caption{HTTPS download/s \lurk{} TLS relative overhead over OpenSSL ($\Delta_{HTTPS}$) in (EC)DHE mode and \dhea{} ($CS_{cert}^{ dhe, r }$). } 
  \label{tab:cs-https-lan}
\end{table}

\subsection{SGX Memory Usage}
The memory that our design needs depends on the chosen configuration. Without session resumption, our implementation uses at most about 25 KB stack and 8 KB heap memory.  Moreover, the memory requirement does not change with the number of connections. Similarly, when session resumption is enabled in the stateless mode, we do not need to store anything in SGX protected memory. However, in the stateful mode, session information needs to be stored in the SGX memory. For each session, we need 104 bytes of SGX memory to store the PSK and session ID. Note that we report memory usage in the debug version (since the Enclave Memory Measurement Tool provided by Intel only works in the debug mode). Therefore, the memory usage will possibly reduce in practice (e.g., due to the optimization in the release mode).

\subsection{\lurk{} TLS Overhead  for HTTPS Over Other Proposals}
\label{sec:https-rw}

In this section we briefly discuss the \lurk{} TLS overhead with the one of other solutions of Section~\ref{sec:relW}, reported in Table~\ref{tab:relw_perf}. 
The comparison is only indicative as the overheads have been made in very different contexts involving different HTTPS servers (NGINX, Apache and wrk) and different TLS libraries (different versions of OpenSSL and libreSSL - though derived from OpenSSL).

\lurk{} is the only design applicable to TLS~1.3 with the lowest overhead in terms of KEX compared to TLS~1.2 and fewest LOC in TEE.
%
As mentioned in Section~\ref{sec:relW}, specific approaches (\lurk{} and Talos) seem to provide a lower overhead over generic frameworks (Graphene -- see Conclave keyless). 
On the other hand, the large overhead measured for Panoply, Graphene, Talos and Conclave-crypt can  be attributed to the resources they require for the protection of the TLS application data. 

In terms of LOC, \lurk{} and Talos limit the potential vulnerability with fewer lines of code and make these solutions more likely to be deployed by cloud providers as less TCB is required. 
This results both in limiting the necessary  memory resources (limited to 90MB) as well as increasing the ability to share the other part of the untrusted library between containers and other applications.   

\begin{table}
    \centering
    \begin{footnotesize}
        \begin{tabular}{|p{0.9cm}|p{1.1cm}|p{0.7cm}|p{0.8cm}|p{0.6cm}|p{0.8cm}|p{1cm}|}
\hline
		&  ${ \scriptstyle \lurk{}}$  & Talos & C-k & C-c & Panoply & Graphene \\
\hline          
{\tiny LOC (K)}      &  3.8     & 5.4 & $>$ 770     &  1300   & 20    & 1300        \\
$ \scriptstyle \Delta_{HTTPS}$  &  0 - 17.7 & 22    & 57               &    81    & 49      & 40 \\
\hline
    \end{tabular}
    \end{footnotesize}
    \caption{HTTPS measurement comparison between Panoply, Graphene, Conclave-keyless (C-k), Conclave-crypt (C-c), Talos. $\Delta_{HTTPS}$ represents the relative overhead (\%) associated to the number of HTTPS download per second for a 1 KB file.}
    \label{tab:relw_perf}
\end{table}

\section{Formal Security Proofs and Analyses}
\label{sec:analysis}

There are two schools of thought w.r.t.\ provable security:
\emph{computational analysis} (a.k.a. \emph{cryptographic proofs}) and \emph{symbolic analysis}.
Computational or provable-security formalisms for security analysis consider messages as bit strings, attackers to be probabilistic polynomial-time algorithms who attempt to subvert cryptographic primitives, and attacks to have a probabilistic dimension of the security parameters.
Computational analysis is proved generally ``by hand'' in a game-based cryptographic model and is appropriated to verify arbitrary corruption and cryptographic AKE (authentication key exchange). 
On the other hand, symbolic models abstract messages to algebraic terms, assume cryptographic primitives to be ideal and not subject to subversion by the adversary, and the attacks be \emph{possibilistic} (i.e., not probabilistic) flaws mounted via a set of Dolev-Yao rules
applied over interleaved protocol executions - assuming cryptography cannot be broken.
Symbolic analysis is tool-assisted, automated, in a protocol-semantics and is appropriated to prove some properties such as PFS for example.


We use both the computational analysis: namely, we extend the (S)ACCE model~\cite{DBLP:conf/eurosp/BhargavanBFOR17} for multi-party AKEs, and we extend the symbolic analysis of a TLS 1.3 draft in ProVerif~\cite{SP2017}. 
We extend the computational analysis to work for the actual current TLS 1.3 and TEEs, as well as applied it to LURK-T. We also extend the symbolic verification to work for TLS 1.3 draft 20 in~\cite{SP2017} (which does not consider some AEAD encrypted payloads during the handshake); then, we apply it to LURK-T.
Importantly, we could forego the computational analysis if there was a computational-soundness result~\cite{DBLP:journals/jar/CortierKW11}, but this does not exist for TLS, let alone for multi-party TLS.

As per Section~\ref{sec:our-protocol}, one can have several modes and several variants of our \lurk\ protocol.
In what follows, we will show security-analyses  for all these variants.
We start by stating  \lurk's requirements semi-formally, in~\ref{subsec:acce-model}. On top of the existing 3(S)ACCE~\cite{DBLP:conf/eurosp/BhargavanBFOR17}  properties, we  add a requirement and a proof for a new property stemming from our use of TEEs; we call this property \emph{trusted key-binding}. 


In Section~\ref{subsec:proofsbyhand}, we provide the computational-security results
for both  \dhea\, and \dhep\ in EC-DHE mode, and discuss in this framework why \dhea\ offers more provable-security guarantees. 
For \dhea\ in EC-DHE mode, if executed in what we call the runtime-attested handshake-context mode, the property we call \emph{trusted key-binding} holds (see Section~\ref{subsec:acce-model}); 
this is a stronger form of accountability (than without the  runtime-attested handshake-context mode), hinging on  TEEs.

In Section~\ref{subsec:proverif}, we use symbolic verification to show that  \dhea\ in EC-DHE mode  attains all the same requirements that TLS1.3 does, and a new, 3-party security property that shows that \client{}, \engine{}, \cs{} have matching views of the handshake even in the presence of a Dolev-Yao attacker.

\subsection{Computational Analysis}

\subsubsection{Cryptographic requirements}
\label{subsec:acce-model}

In order to give our cryptographic proofs that \lurk\ achieves its security goals, we
  use the recent 3(S)ACCE formal security model for proxied AKE~\cite{DBLP:conf/eurosp/BhargavanBFOR17}.

 In  essence, we will use this 3(S)ACCE model, extended with an additional 4th party, namely the \emph{attester \attester}, who interacts  with the \cs{} and (the \attester\  may be called upon via $\engine{}$), but this interaction \attester-\cs{} is outside of the ACCE computation. Because of this, we continue to call the model 3(S)ACCE (as in, with 3 parties); we just make a note that a 4th party -- the attester-- is present, ``out of band''.

\subhead{Security requirements for \lurk}
For \lurk, we 
prove the  {3(S)ACCE}  requirements: i.e., entity authentication, channel security and accountability.
Below, we add a new one, linked to the attester party, and call this requirement \emph{trusted key-binding}.

\subhead{\lurk\ with runtime-attested handshake-context} 
In this sub-variant, a \emph{runtime} TEE system is called to yield a separate ``quote'' over the whole handshake done inside the \cs{} during a TLS session.
 So, we request the quote from the remote enclave (found on the \cs{}) and verify this using the  Attestation Service. Namely, we request the quote as soon as the \cs{} prepares the \ssign{} and before it does so.
 Then, we encrypt the buffer containing the operations on the \cs{} and its arguments (it will just contain $\handshake $), with the shared key established via remote attestation  (e.g., $seal\_{key}$). In this optional sub-variant, this step is done and the Attestation Service therefore will receive a ``binding''/``context'' to the channel-key calculated during the handshake.
We call this type of \lurk\  -- \emph{\lurk\ with runtime-attested handshake-context.}

\subhead{Trusted key-binding} We now state our new attestation-relevant property more widely than for \lurk{}, for a server-controlled delegated TLS achieves trusted key-binding with runtime attestation on the \cs{}. We say a \emph{server-controlled delegated TLS achieves trusted key-binding} if  \cs{}
is  able to compute the channel keys $\ck$ used by \client{} and \engine{} and the  handshake context/transcript corresponding to  these keys $\ck$  is  asynchronously attested. 
That is, if presented with this handshake context  by the attester again, then \cs{} can recompute these keys $\ck$ and produce the same $\ssign, h, a, etc$ sent to \engine{} in the handshake where the keys ($\ck$) were used.

Note that the attester gets only necessary parameters from the handshake.
Notably, it does not get $\ssign{}, h, a$, so it cannot impersonate the \cs{} or resume a session as a \cs{}. Further, in some TEE systems (e.g., if we use a TPM -- trusted platform module), we could open a ``TEE session'' for the whole part of the handshake run on the \cs{} and sign that as a  proof of computation for the attester, yet we deliberately go against that. Such a design would give the attester all the information of the computation on the \cs{} side which we believe will  place too much trust on the attester, allowing it to see \emph{long-term} secrets of the \cs{} pertaining to another, specific protocol, i.e., TLS.

Finally,  trusted key-binding  is a type of enhanced 3(S)ACCE accountability which is based on the \lurk\ \cs{} executing its part of the TLS-server in an enclave.

\subsubsection{Cryptogtaphic proofs}
\label{subsec:proofsbyhand}

W.r.t.\ the properties recalled/given above, we now state our cryptographic guarantees.

\subhead{Entity-authentication result}
\emph{If TLS~1.3 is secure w.r.t.\  unilateral entity authentication, if the protocol between \engine{} and \cs{} is
a secure ACCE protocol or they are collocated, if the two protocols (the one
between \client{} and \engine{}, and the one between \engine{} and \cs{}) ensure 3(S)ACCE mixed entity 
authentication~\cite{DBLP:conf/eurosp/BhargavanBFOR17} in the case where \engine{} and \cs{} are not collocated, 
if the signature and hash in TLS~1.3 server-side are secure in their respective threat models,
if the authentication encryption used in TLS~1.3 is secure in its model, then \dhea\ and \dhep\ in EC-DHE mode are entity-authentication secure in the  3(S)ACCE model.}
%

\subhead{Channel security result} 
\emph{If TLS~1.3 is secure w.r.t.\  unilateral entity authentication, if the protocol between \engine{} and \cs{} is
a secure ACCE protocol or they are collocated, if the two protocols ensure  3(S)ACCE mixed entity authentication~\cite{DBLP:conf/eurosp/BhargavanBFOR17} in the case where \engine{} and \cs{} are not collocated, if the signature in TLS~1.3 server-side  is secure in its threat model,  if the authentication encryption used in TLS~1.3 is secure in its model, and the freshness function is a non-programmable PRF~\cite{DBLP:conf/latincrypt/BoureanuMV12}, then  \dhep\ in EC-DHE mode are entity-authentication secure in the  3(S)ACCE model attain channel security in the  3(S)ACCE model.}
%

Note that the two security results above apply to all variants and sub-variants of \lurk{}. 
These two requirements are the main requirements for any AKE protocol, now cast and proven here not over two but over three parties, in the 3(S)ACCE model. 
This alone makes \lurk{} a secure TLS decoupling between the Crypto Service to the Engine. 
So, the next two statements can be viewed as ``bonus'' security, attained only by the  variants of \lurk{} which are computationally more expensive.

\subhead{Accountability result}
\emph{If TLS~1.3 is secure w.r.t.\  unilateral entity authentication, if the protocol between \engine{} and \cs{}  is
a secure ACCE protocol  or they are collocated, if the two protocols ensure  3(S)ACCE mixed entity authentication in the case where \engine{} and \cs{} are not collocated,  and the freshness function is a non-programmable PRF~\cite{DBLP:conf/latincrypt/BoureanuMV12}, then \dhea\   attains accountability in the  3(S)ACCE model.}
%

Accountability requires that \cs{} always be able to compute all keys and sub-keys of the session established between the client and \engine{}.  So, accountability is incompatible when session-resumption is done by the \engine{} alone (i.e., \dhep). That is the above security statement w.r.t.\ accountability only holds for \dhea.
Note that this is not critical in practice. 
Also, it comes at a cost (i.e., \dhea\ is more computationally expensive expensive than \dhep). 
So, with \lurk{}, we provide a series of variants, allowing the deployment-stage to choose between security  and efficiency.

\subhead{Trusted key-binding result}
\emph{If TLS~1.3 is secure w.r.t.\  unilateral entity authentication, if the protocol between \engine{} and \cs{}  is
a secure ACCE protocol or they are collocated, if the two protocols ensure  3(S)ACCE mixed entity authentication in the case where \engine{} and \cs{} are not collocated,  and the freshness function is a non-programmable PRF~\cite{DBLP:conf/latincrypt/BoureanuMV12} and if the TEE allows for runtime remote attestation, then \dhea\   attains trusted key-binding.} 

Trusted key-binding can be seen as an attested form of accountability. So, like accountability, it will only hold for variants of \lurk\ where there is no session resumption by the \engine{} alone. 
Again, this is not critical in practice -- since trusted key-binding is an arguably very strong  requirement of security and trust.

\subsection{Symbolic Verification}
\label{subsec:proverif}
We perform a symbolic verification using ProVerif~\cite{DBLP:journals/ftsec/Blanchet16} to show that the \lurk\ protocol, from a symbolic verification perspective, attains the same security properties as TLS~1.3, along with additional properties as described below.
In this section, we focus on the verification of \dhea.
We first  show that \lurk\ does not impact TLS security (from a symbolic-verification perspective). Then, we show that the addition of the 3rd party still attains security w.r.t.\ symbolic verification. All this is complementary to the results in Section~\ref{subsec:proofsbyhand}.

This section is structured as follows. First, we report on a ProVerif-verification of TLS~1.3 which we lifted from 
TLS~1.3 pre-standardisation (i.e., draft 18) to the current standard. Then, we show that all the 2-party, TLS~1.3-centred properties are preserved on \lurk. We also add a new 3-party agreement property for \lurk{}, which  ProVerif proves to hold, thus showing \lurk{} to be a secure proxied TLS. All our ProVerif files and  results are available at: \url{https://github.com/lurk-t/proverif}.

\subsubsection{Verifying standardised TLS~1.3  in ProVerif}
\label{sec:tlsproverif}
Our approach was to reuse a ProVerif specification of a draft of TLS~1.3, given in~\cite{prosecco,SP2017}. The latest available version of this specification encoded draft 20 of TLS~1.3 pre-standardisation (no newer version as confirmed by the authors). 
%
So, first, we  updated this existing ProVerif specification of TLS with  the RFC 8446. In short, the ProVerif  model did not specify the handshake to include AEAD encryption for the \textit{Certificate}, \textit{CertificateVerify} and \textit{Finished} messages.  We applied the necessary updates to ProVerif models  and verified that the original properties still held. The only difference observed is that, in our newly updated models for standard-TLS~1.3,   the automatic proofs take longer, as we detail below.

\subsubsection{Verifying \lurk\  in ProVerif}
\label{sec:lurkproverif}

We modelled \lurk\ in ProVerif. We therefore split the ProVerif  \server{}-process in two: a \cs{} process and an \engine{} process. In each, we  encoded \dhea\  and, specifically, also the case in  which the \cs{} and \engine{} are not collocated. In this case, we modelled a secure channel between the \cs{} and  \engine{}, as per the \lurk\ specifications; in ProVerif, this is what is called a private channel, not accessible to the underlying Dolev-Yao attacker. We inherited all the Diffie-Hellman exponentiation aspects (including modelling weak subgroups) from the TLS implementation.
Note that we do not model the TEE specifically, but since \cs{} cannot be adaptively corrupted in the model at hand (which is the case symbolic verification), that equates to the TEE being modelled ``by default''.

The query we added to the ones inherited from TLS~1.3 expresses that there is always a correct/secure session interleaving and execution between the \client{}, \engine{} and \cs{}, even with the Dolev-Yao attacker in the middle. In practice, this means that a Dolev-Yao attacker cannot find a way to mis-align the execution of the three parties by doing a man-in-the-middle-type attack.

\begin{figure}
\begin{tiny}
\begin{verbatim}
query cr:random, sr:random, cr':random, sr':random, 
      psk:preSharedKey,p:pubkey, e:element,
      o:params, m:params,  
      ck:ae_key,sk:ae_key,ms:bitstring,cb:bitstring, log:bitstring; 
      
      inj-event(ClientFinished(TLS13,cr,sr,psk,p,o,m,ck,sk,cb,ms)) ==>
      (inj-event(PreServerFinished(TLS13,cr,sr,psk,p,o,m,ck,sk,cb))  ==>
		(inj-event (TLS13_recvd_CV (cr, sr, p, log))  ==>
			(inj-event (CS_sent_CV(cr, sr, p, log) ) ==>
				inj-event (TLS13_sent_cr_sr_to_CS (cr, sr, p, log)) 
			)
		)
	  )
      || (event(WeakOrCompromisedKey(p)) &&  (psk = NoPSK 
			|| event(CompromisedPreSharedKey(psk)))) ||  
      event(ServerChoosesKEX(cr,sr,p,TLS13,DHE_13(WeakDH,e))) ||    
      event(ServerChoosesHash(cr',sr',p,TLS13,WeakHash)).   
\end{verbatim}
\end{tiny}
\caption{Agreement query between \client{}, \engine{}, \cs{} }
  \vspace*{-0.2cm}
\label{fig:proverif-property-1}
\end{figure}

As shown in Fig.~\ref{fig:proverif-property-1}, our added query  captures the execution of the following sequence of events: 1) \begin{small}TLS13\_sent\_cr\_sr\_to\_CS \end{small}\normalfont denoting that \engine{} contacted the \cs{} with clinet{}’s handshake details; 2) \begin{small}CS\_sent\_CV \end{small}\normalfont  denoting that the CryptoService sent a signed share to \engine{}; 3) \begin{small}TLS13\_recvd\_CV \end{small}\normalfont  denoting that  \engine{} got from the said share signed from the CryptoService; 4) \begin{small}PreServerFinished \end{small}\normalfont  denoting that \engine{} acted as a TLS server and reached the point of sending out a DH share to the client; 5) \begin{small}ClientFinished \end{small}\normalfont  denoting that \client{} finished a handshake.
By considering the introduced parameters in these events, one can observe that the data is bound among all such events during any given execution. These events are required to be injective, implying a one-to-one mapping in occurrences between them. Therefore, not only must this sequential events and data agreement hold for every \lurk{} execution, but each \cs{} execution will also uniquely correspond to a single \engine{} execution and a single \client{} execution, through a distinct set of matching handshake data. This security agreement goal is demonstrated to persist, except in the cases of compromised \cs{}, or compromised PSK, or due to the use of a weak DH subgroup, or a weak hash function.
Such exceptions are comprehensively addressed by the list of disjunct terms appended at the end of the query.
\subsubsection{Experimental setup}
We conducted our ProVerif  verification in two settings: 
(a) using an Ubuntu 20.04 Focal VM with a V100 GPU and 128 GB RAM on KVM;
(b) using a laptop with Windows 10 and Intel(R) Core(TM)
i7-8650U CPU @1.90GHz and 32 GB RAM and the latest version 2.02 for ProVerif. 
 While the powerful setting (a) is evidently very suitable for the development phase of our ProVerif models and for fast verification, we consider that setting (b) is more plausible to be used for reproducing our proofs. In setting (b), without the option to generate the attack graphs, analyzing all 29 queries automatically (the 24 queries inherited from \cite{prosecco} plus the 5 queries we added specifically for \lurk{} including the query detailed above), takes 17.25 hours. Verified separately, the query in Figure~\ref{fig:proverif-property-1} requires 1.5 hours to be proved (true).

\section{Potential Vulnerabilities}
Although \lurk\ protocol's security goals are formally verified, a \lurk{} deployment may still face practical security challenges. For example, several new SGX attacks have been reported almost every year for the past several years---see e.g., the recent Downfall~\cite{sgx-attack-usenix23} and {\AE}PIC attacks~\cite{sgx-attack-usenix22}; a survey of attacks, CPU update delays, and exploits for commercial SGX applications are provided by \url{SGX.fail}~\cite{sgxfail}; and a comprehensive list\footnote{\url{https://www.intel.com/content/www/us/en/developer/topic-technology/software-security-guidance/processors-affected-consolidated-product-cpu-model.html}} is also maintained by Intel. A \lurk{} deployment must remain up-to-date with all pertinent fixes (SDK and microcode, if available). Missing any such updates, or attacks with no available mitigations may make any SGX deployment vulnerable, including \lurk{}. Possible vulnerabilities may also be introduced in \lurk{}'s implementation. Our code is open-sourced, and relatively small in size (under 4K LOC), and thus the possibility of such vulnerabilities is relatively low. The private keys that \lurk{} aim to protect, can still be leaked through content owner's negligence; e.g., if keys are provisioned without attesting a \cs{} implementation (solutions such as Blindfold~\cite{Blindfold} can be used for secure key provisioning).

\section{Conclusions}
\label{sec:concl}

We introduced  \lurk{} -- a provably secure and  efficient extension of TLS 1.3, and of the generic LURK framework.
We designed \lurk{} with a TLS server  decoupled into a \emph{\lurk{} TLS Engine}  and a \emph{\lurk{} Crypto Service}  and split  the TLS handshake across the two modules; the Crypto Service is executed inside a TEE and it accepts very specific and limited requests. We offered several modular variants of \lurk{}, balancing security and efficiency. 
In addition, we implemented the Crypto Service using Intel SGX, and integrated our implementation to OpenSSL with minor changes. Finally, our experimental results looked at \lurk's overheads compared to TLS 1.3 handshakes and demonstrated that it provides competitive efficiency.

\ifthenelse{\boolean{acm}}{
  \bibliographystyle{ACM-Reference-Format}
}{
  \bibliographystyle{IEEEtranS}
}
%


\begin{IEEEbiography}[{\includegraphics[width=1in,height=1.2in,clip,keepaspectratio]{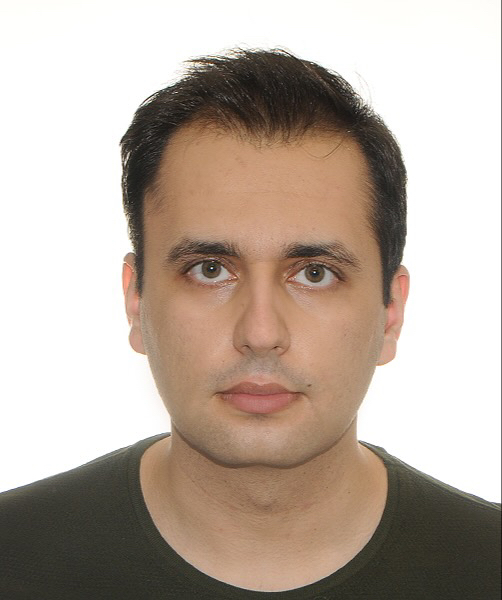}}]{Behnam Shobiri}  is a security researcher at Tigera wher he is researching Cloud and Kubernetes security. Prior to that, he was a master's student at Concordia University and worked on TLS and CDN security. He got his bachelor's degree from the Ferdowsi University of Mashhad in the field of computer engineering. 
\end{IEEEbiography}

\begin{IEEEbiography}[{\includegraphics[width=1in,height=1.2in,clip,keepaspectratio]{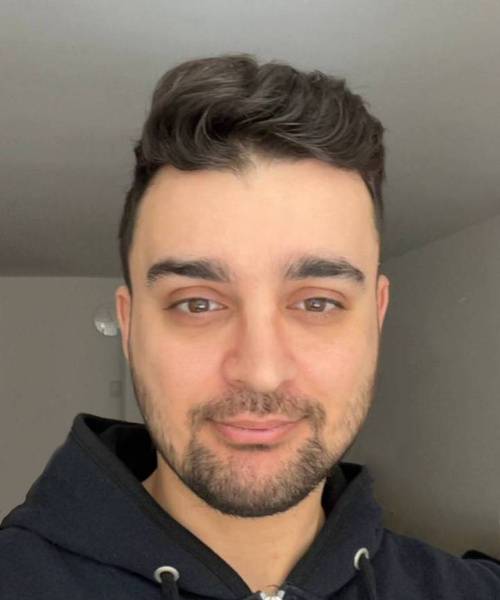}}] {Sajjad Pourali} is currently pursuing his Ph.D. degree in Information and Systems Engineering at Concordia University. His research interests include internet, application and system security and privacy.
\end{IEEEbiography}

\vspace{-1cm}

\begin{IEEEbiography}[{\includegraphics[width=1in,height=1.2in,clip,keepaspectratio]{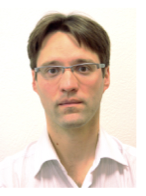}}]
{Daniel Migault}  is an expert in the Ericsson cybersecurity team and is actively involved in standardizing security protocols at the IETF.
\end{IEEEbiography}

\vspace{-1cm}
\begin{IEEEbiography}[{\includegraphics[width=1in,height=1.2in,clip,keepaspectratio]{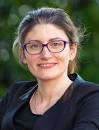}}]{Ioana Boureanu}   is a Professor in Secure Systems at University of Surrey.  She is the deputy director of Surrey Centre for Cyber Security, the co-director of University of Surrey Gold-level ACE-CSE, as well as Director of our GCHQ-accredited Information Security MSc.

\end{IEEEbiography}

\vspace{-1cm}

\begin{IEEEbiography}[{\includegraphics[width=1in,height=1.2in,clip,keepaspectratio]{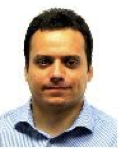}}]
{Stere Preda}  received his PhD in Computer Science from TELECOM Bretagne, France. He is currently a senior researcher with expertise in cybersecurity at Ericsson. He has been an  an active contributor to ETSI NFV security standardization.
\end{IEEEbiography}

\vspace{-1cm}
\begin{IEEEbiography}[{\includegraphics[width=1in,height=1.2in,clip,keepaspectratio]{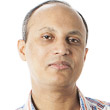}}]
{Mohammad Mannan} 
in an associate professor at the Concordia Institute for Information Systems Engineering, Concordia University.  His research interests lie in the area of Internet and systems security. Dr. Mannan is involved with several well-known conferences (e.g., USENIX Security, ACM CCS), and journals (e.g., IEEE TIFS and TDSC).

\end{IEEEbiography}
\vspace{-1cm}
\begin{IEEEbiography}[{\includegraphics[width=1in,height=1.2in,clip,keepaspectratio]{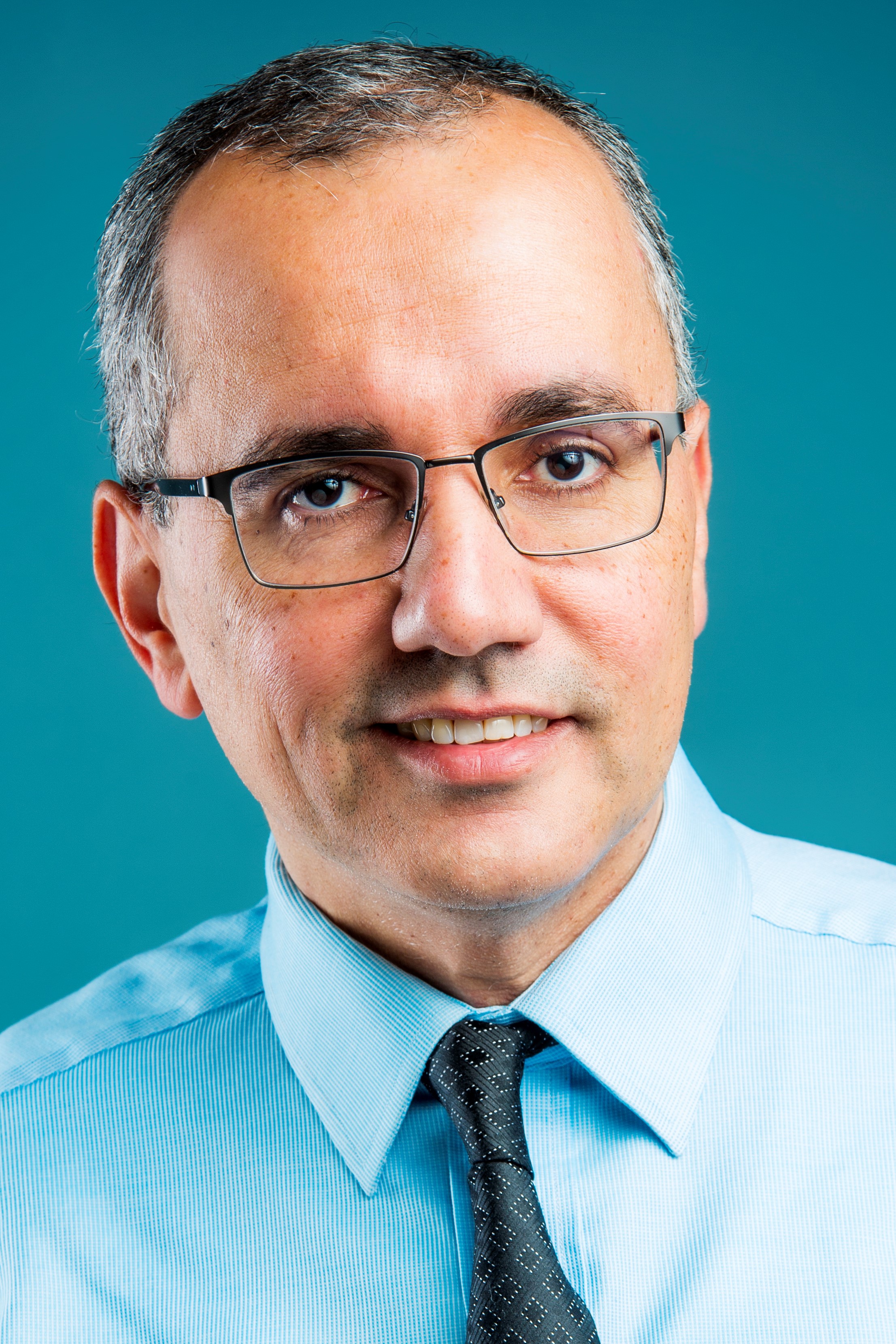}}]
{Amr Youssef} received the B.Sc.\ and M.Sc.\ degrees from Cairo University, Cairo, Egypt, in 1990 and 1993 respectively, and the Ph.D. degree from Queens University, Kingston, ON., Canada, in 1997. Dr.\ Youssef is currently a professor at the Concordia Institute for Information Systems Engineering (CIISE) at Concordia University, Montreal, Canada.
His research interests include cryptology, security and privacy. 
\end{IEEEbiography}

\end{document}